\documentclass[aip,12pt]{revtex4-2}

\usepackage{amsmath,bm}
\usepackage{amssymb}
\usepackage{mathrsfs}
\usepackage{amsfonts}
\usepackage{graphicx}
\usepackage{siunitx}
\usepackage[version=3]{mhchem}
\usepackage[acronym]{glossaries}

\DeclareSIUnit\au{\text {au}}

\newcommand{\ke}[1]{\vert #1 \rangle}

\newacronym{cboa}{CBOA}{cavity Born-Oppenheimer approximation}
\newacronym{bo}{BOA}{Born-Oppenheimer approximation}
\newacronym{cbohf}{CBO-HF}{cavity Born-Oppenheimer Hartree-Fock}
\newacronym{scf}{SCF}{self-consistent field}
\newacronym{dse}{DSE}{dipole self-energy}
\newacronym{pes}{PES}{potential energy surface}
\newacronym{cpes}{cPES}{cavity potential energy surface}
\newacronym{vsc}{VSC}{vibrational-strong coupling}
\newacronym{esc}{ESC}{electronic-strong coupling}
\newacronym{lp}{LP}{lower polariton}
\newacronym{up}{UP}{upper polariton}
\newacronym{etc}{ETC}{extended Tavis-Cummings}
\newacronym{dqc}{DQC}{double quantum coherence}
\begin{document}

\title{Disentangling collective coupling in vibrational polaritons with double quantum coherence spectroscopy}

\author{Thomas Schnappinger}
\email{thomas.schnappinger@fysik.su.se}
\affiliation{Department of Physics, Stockholm University, AlbaNova University Center, SE-106 91 Stockholm, Sweden}
\author{Cyril Falvo}
\affiliation{Universit\'e Paris-Saclay, CNRS, Institut des Sciences Mol\'eculaires d’Orsay, 91405 Orsay, France}
\affiliation{Universit\'e Grenoble-Alpes, CNRS, LIPhy, 38000 Grenoble, France}
\author{Markus Kowalewski}
\email{markus.kowalewski@fysik.su.se}
\affiliation{Department of Physics, Stockholm University, AlbaNova University Center, SE-106 91 Stockholm, Sweden}

\date{\today}%

\begin{abstract}
Vibrational polaritons are formed by strong coupling of molecular vibrations and photon modes in an optical cavity. 
Experiments have demonstrated that vibrational strong coupling can change molecular properties and even affect chemical reactivity.
However, the interactions in a molecular ensemble are complex, and the exact mechanisms
that lead to modifications are not fully understood yet.
We simulate two-dimensional infrared spectra of molecular vibrational polaritons based on the double quantum coherence technique to gain further insight into the complex many-body structure of
these hybrid light-matter states.
Double quantum coherence uniquely resolves the excitation of hybrid light-matter polaritons and allows to directly probe the anharmonicities of the resulting states.
By combining the cavity Born-Oppenheimer Hartree-Fock ansatz with a full quantum dynamics simulation of the corresponding eigenstates, we go beyond simplified model systems. 
This allows us to study the influence of self-polarization and the response of the electronic structure to the cavity interaction on the spectral features even beyond the single-molecule case.  
\end{abstract}

\maketitle
\section{Introduction}

Light-matter coupling between optical resonances of a cavity and molecular transitions can result in the formation of molecular polaritons~\cite{Basov2021-hr,Dunkelberger2022-oh,Ebbesen2023-fd,Bhuyan2023-se}. 
When coupling overcomes the dissipative processes, the strong coupling regime is reached, and these new hybridized states with mixed photon-matter character can be observed spectroscopically.
A pair of characteristic peaks, called the \gls{lp} and \gls{up}, can be observed
shifted above and below the field-free transition frequency.
By controlling of the photonic environment it becomes possible to selectively couple the cavity photon modes to vibrational or electronic transitions in molecules, called \gls{vsc} and \gls{esc}, respectively.
Both types of strong coupling are discussed as effective tools not only for modifying photophysics and photochemistry~\cite{Hutchison2012-od,Munkhbat2018-rh,Peters2019-vt,Timmer2023-bp} but also for altering electronic ground-state reaction rates and product branching ratios~\cite{Thomas2019-ve,Bai2023-ev,Ahn2023-qk}.
However, the current theoretical description of \gls{vsc} in particular is far from complete and many open questions remain. For example, it is not clear how the polariton states contribute to the
experimentally observed modification of reaction rates.
The formation of polaritons in an ensemble is inherently a collective phenomenon in which
the excitation is delocalized over the whole system.
However, a chemical reaction is thought to occur locally in individual molecules. 

Consequently, both linear~\cite{DelPo2020-kq,Avramenko2021-hs,Hirschmann2024-ym} and non-linear spectroscopy~\cite{Xiang2018-vn,Mewes2020-qv,Autry2020-wy,Son2022-sg,Quenzel2022-uh,Mondal2023-fl,Ren2024-zs,Sufrin2024-ue} are methods that can be used to advance the understanding of molecular polaritonics.
In particular, coherent multidimensional infrared spectroscopy is a powerful tool that can probe anharmonicities and vibrational energy relaxation pathways, opening the possibility of further insight into processes under \gls{vsc}.
Among the multipulse non-linear spectroscopic techniques available~\cite{Mukamel1995-wa} we focus on the \gls{dqc} technique~\cite{Cervetto2004-zg,Fulmer2004-jv,Abramavicius2008-gu,Saurabh2016-tc,Sufrin2024-ue} in this work. 
The study of \gls{dqc} for molecular vibrational polaritons in optical cavities allows one to directly probe the double excitation manifold without being convolved with single exciton resonances.
Furthermore, the effect of strong couplings on the vibrational anharmonicities in the systems can be observed. 

In this manuscript we simulate \gls{dqc} spectra for a single \ce{HF} molecule and a pair of \ce{HF} molecules resonantly coupled to a single-photon mode of an optical cavity in the infrared. 
Our simulation is based on full-dimensional \glspl{cpes} calculated on the \gls{cbohf}~\cite{Schnappinger2023-hh,Sidler2024-vm} level of theory. 
The \gls{cbohf} ansatz allows the molecular system, e.g., the electronic structure, to respond to the cavity field due to the \gls{scf} procedure, and the inclusion of \gls{dse} terms allows self-polarization~\cite{Schafer2020-cb,Sidler2022-cg,Schnappinger2023-hh,Sidler2024-vm} of the coupled cavity-molecular system. 
In addition to a detailed analysis of the obtained \gls{dqc} spectra, we also investigated the influence of the \gls{dse}, e.g. self-polarization, and the \gls{scf} procedure, e.g. response of the electronic structure, on the spectral features. 
Finally, we study the changes introduced by going beyond a single molecule and the possible influence of cavity-mediated intermolecular interactions on the \gls{dqc} signal.

\section{Theory and models}
\subsection{\label{sec:cboa} Cavity Born-Oppenheimer approximation}
In our recent work~\cite{Schnappinger2023-hh,Schnappinger2023-wp} we have introduced the \gls{cbohf} approach, which represents a formulation of the well-known Hartree-Fock ansatz in the context of \gls{cboa}. 
Within \gls{cboa}, the cavity field modes are grouped with the nuclei in a generalized Born-Huang expansion~\cite{Schafer2018-vf,Ruggenthaler2023-aa}, and then one can subsequently solve the quantum problem of the electrons and then of the nuclei and photons. 
The electronic \gls{cboa} Hamiltonian for a single cavity field mode takes the form of
\begin{equation}
\label{eq:h_cbo}
\hat{H}_{CBO} = \hat{H}_{el}  + \frac{1}{2} \omega_c^2 q_{c}^2 - \omega_c q_{c} \left(\bm{\lambda}_{c} \cdot \bm{\hat{\mu}} \right) + \frac{1}{2} \left(\bm{\lambda}_{c} \cdot \bm{\hat{\mu}} \right)^2  \,, 
\end{equation}
where $\bm{\hat{\mu}}$ represents the molecular dipole operator and $\hat{H}_{el}$ is the Hamiltonian for the field-free many-electron system. 
The second term is a harmonic potential introduced by the photon displacement field, with the classic photon displacement coordinate $q_{c}$ and $\omega_c$ being the frequency of the cavity mode.
The third term describes the dipole coupling between the molecular system and the photon displacement field, which is characterized by the coupling strength $\bm{\lambda}_{c}$.
The last term is the \gls{dse} operator~\cite{Rokaj2018-ww,Schafer2020-cb,Sidler2022-cg}, which describes the self-polarization of the molecule-cavity system.
The cavity mode-specific coupling parameter $\bm{\lambda}_{c}$ for a cavity with effective mode volume $V_c$ is defined as follows:
 \begin{equation}
 \label{eq:lam}
\bm{\lambda}_{c} =  \bm{e}_{c} \lambda_{c} =  \bm{e}_{c}  \sqrt{\frac{4 \pi}{V_c}}\,.
\end{equation}
The unit vector $\bm{e}_{c}$ denotes the polarization axis of the cavity mode. 
In this work, we make use of two variations of the electronic \gls{cboa} Hamiltonian $\hat{H}_{CBO} $, describing the many-electron problem, and solving both cases in the \gls{scf} \gls{cbohf} ansatz~\cite{Schnappinger2023-hh,Schnappinger2023-wp}.  
In the first case labeled linear \gls{cbohf} we neglected the \gls{dse} operator and the resulting energy $E_{CBO}^{lin}$ reads:
\begin{equation}
\label{eq:e_cbo_lin}
\bigl\langle  E_{CBO}^{lin} \bigr\rangle  = E_{el} +  E_{lin} + E_{dis} \quad  \text{ with} \quad E_{dis} = \frac{1}{2} \omega_c^2 q_{c}^2\,.
\end{equation}
For the second case, the standard \gls{cbohf} Hamiltonian including the \gls{dse} operator is used to self-consistently determine the energy $E_{CBO}$:
\begin{equation}
\label{eq:e_cbo_full}
\bigl\langle  E_{CBO} \bigr\rangle  = E_{el} +  E_{lin} + E_{dse} + E_{dis} \,.
\end{equation}
In addition to considering the full \gls{dse} operator, the \gls{scf} treatment itself has proved crucial to capture relevant aspects in the description of strongly coupled molecules~\cite{Sidler2024-vm,Schnappinger2023-hh,Schnappinger2023-wp}. 
To study this aspect, we used the field-free Hartree-Fock energy and expectation values of the dipole moment and the \gls{dse} operator to construct a \gls{etc} model~\cite{Jaynes1963-os,Kowalewski2016-zo,Gudem2021-um,Schnappinger2023-ar,Borges2024-pn}.
The corresponding energy expectation value that defines the \glspl{cpes} has the following form 
\begin{equation}
\label{eq:Hetc}
\bigl\langle  E_{ETC} \bigr\rangle  = E_{el} - \omega_c q_{c}  \bigl\langle\bm{\lambda}_{c} \cdot \bm{\hat{\mu}} \bigr\rangle_0 + \frac{1}{2} \bigl\langle \left(\bm{\lambda}_{c} \cdot \bm{\hat{\mu}} \right)^2 \bigr\rangle_0 + \frac{1}{2} \omega_c^2 q_{c}^2
\end{equation}
Both the linear light-matter interaction term, as well as a quadratic \gls{dse} term are calculated with the corresponding field-free expectation values $\bigl\langle\bm{\lambda}_{c} \cdot \bm{\hat{\mu}} \bigr\rangle_0$ and $\bigl\langle \left(\bm{\lambda}_{c} \cdot \bm{\hat{\mu}} \right)^2 \bigr\rangle_0$ respectively.

\subsection{Double Quantum Coherence Spectroscopy}

\begin{figure}
     \centering
    \includegraphics[width=0.5\textwidth]{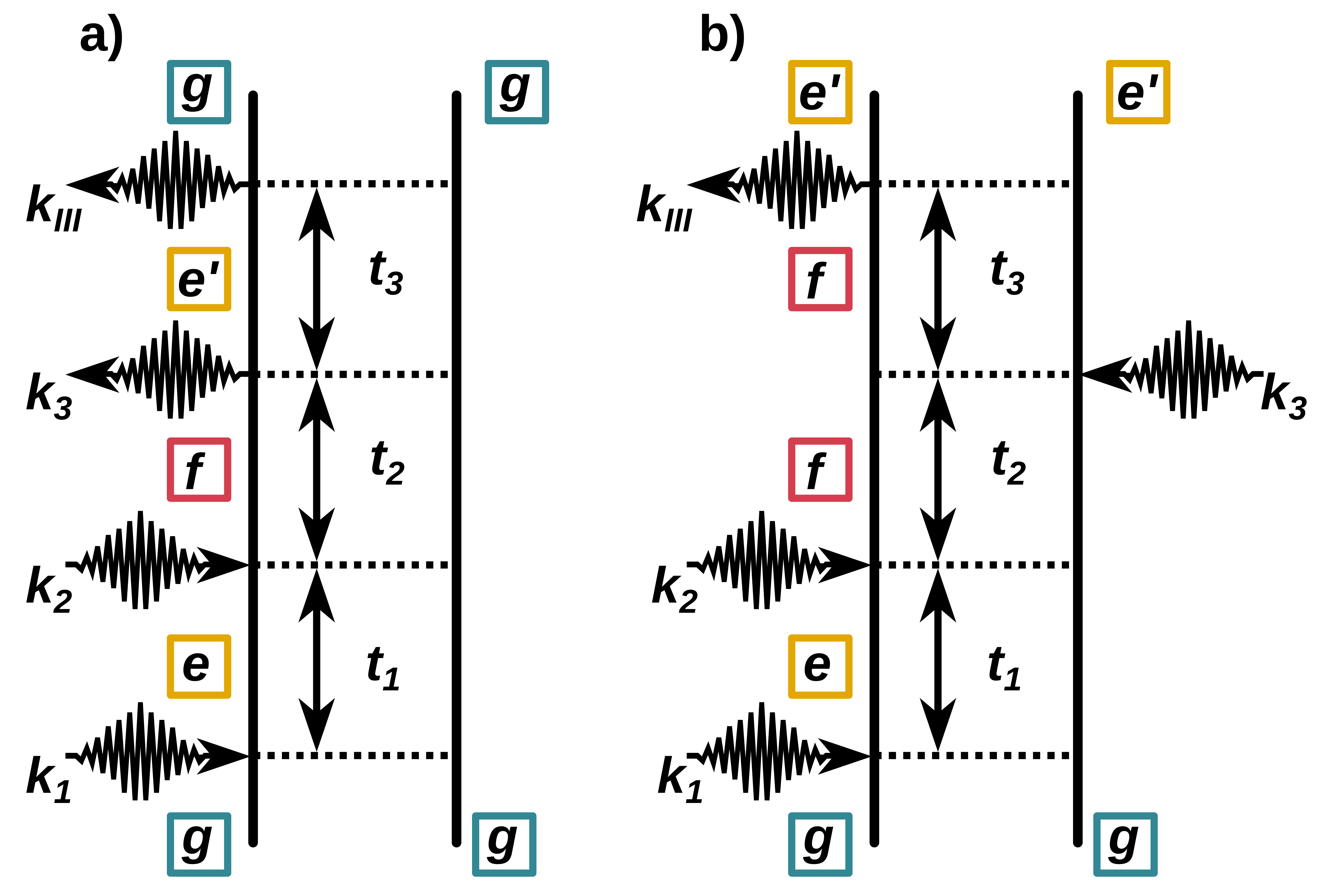}
    \caption{The two double sided Feynman diagrams that contribute to the \gls{dqc} signal $\mathcal{S}(t_3, t_2, t_1)$ in the phase-matching direction ($\bm{k}_{III} = \bm{k}_1 + \bm{k}_2 - \bm{k}_3$) in the rotating wave approximation. The ground state is colored green, the single excited states yellow, and the double excited states are red.}%\CF{should we specify e or e' in the diagram to connect it ot Eq.6 ?} 
\label{fig:dia}
\end{figure}

The \gls{dqc} technique is a 2D-IR method performed with four temporally well-separated laser pulses.
The first three pulses with wavevectors $\bm{k}_1$, $\bm{k}_2$, and $\bm{k}_3$ generate a non-linear polarization in the molecular ensemble, which is heterodyne detected with the fourth pulse. We focus on the signal generated along the phase matching direction $\bm{k}_{III} = \bm{k}_1 + \bm{k}_2 - \bm{k}_3$~\cite{Mukamel1995-wa,Abramavicius2008-gu,Saurabh2016-tc}. 
The signal recorded against the three delay times between the pulses $t_1$, $t_2$,
and $t_3$ is denoted by $\mathcal{S}(t_3, t_2, t_1)$.
By invoking the rotating wave approximation, only the dominant contributions to $\mathcal{S}$, where all interactions are resonant, are retained.
Consequently, there are only two contributions to the $\bm{k}_{III}$ signal for our coupled cavity molecule systems represented by the double-sided Feynman diagrams shown in Fig.~\ref{fig:dia}.
The two diagrams represent the same evolution during the first two time intervals: during $t_1$ the density matrix oscillates with the frequency $\Omega_{eg} =\epsilon_e - \epsilon_g$, and during $t_2$ the density matrix oscillates with the frequency $\Omega_{fg} =\epsilon_f - \epsilon_g$. 
Here $\epsilon_g$, $\epsilon_e$ and $\epsilon_f$ are the vibrational/polaritonic eigenenergies of the vibrational/polaritonic ground state $g$ (green), the first-excitation manifold $e$ (yellow) and the second-excitation manifold $f$ (red).
During $t_3$ the diagrams yield an oscillation frequency of either $\Omega_{e'g}$, or $\Omega_{fe'}$ (see Figs.~\ref{fig:dia}~a) and ~\ref{fig:dia}~b) respectively). 
Thus, transitions that form a harmonic ladder, that is, $\Omega_{e'g}=\Omega_{fe'}$, cancel, and such transitions vanish in the \gls{dqc} signal.
Consequently, one of the main features of the \gls{dqc} technique is its sensitivity to anharmonicities in the molecular systems studied or, in our case, polaritonic systems. 
Another one is the possibility of having direct access to information on not only the first but also on the second excitation manifold.
The obtained 3D signal $\mathcal{S}(t_3, t_2, t_1)$ can be written as double Fourier transform with respect to $t_3$ and $t_2$:
\begin{equation}
\label{eq:DQC1}
\mathcal{S}\left(\Omega_3,\Omega_2,t_1\right) = \int_{0}^{\infty} \int_{0}^{\infty} dt_3 dt_2 e^{i\left(\Omega_3t_3 + \Omega_2t_2\right)} \mathcal{S}\left(t_3,t_2,t_1\right)\,.
\end{equation}
Expanding the expression $\mathcal{S}\left(\Omega_3,\Omega_2,t_1\right)$ in the basis of the polaritonic eigenstates, for $t_1 = 0$, the signal becomes
\begin{equation}
\label{eq:DQC2}
\mathcal{S}\left(\Omega_3,\Omega_2, t_1=0 \right)  = \sum_{e,e',f} \frac{1}{\left( \Omega_2 -\Omega_{fg} + i \gamma\right)}  \times \left[ \frac{\mu_{ge'} \mu_{e'f}  \mu_{fe} \mu_{eg}}{\left( \Omega_3 -\Omega_{e'g} + i \gamma\right)} - \frac{\mu_{ge'} \mu_{e'f} \mu_{fe} \mu_{eg}}{\left( \Omega_3 -\Omega_{fe'} + i \gamma\right)} \right]\,,
\end{equation}
where $\mu_{ij}$ are the transition dipole moments between polaritonic states $i \rightarrow j$  and $\gamma$ is an empirical dephasing rate.

\subsection{Computational Details}

The vibrational eigenenergies and the transition dipole moments needed to calculate the \gls{dqc} spectra according to Eq.~\ref{eq:DQC2} are obtained using the \gls{cbohf} ansatz, in the Psi4NumPy environment~\cite{Smith2018-tu}, which is an extension of the PSI4~\cite{Smith2020-kq} electronic structure package. 
All calculations were performed using the aug-cc-pVDZ basis set~\cite{Kendall1992-wu} and the geometry of the isolated single \ce{HF} molecule was optimized at the Hartree-Fock level of theory.
In this study, we will consider the interaction between a single lossless mode cavity and an ensemble of $N_{mol}=1$ and $N_{mol}=2$ \ce{HF} molecules. We assume that the molecules are aligned with the polarization of the cavity mode and we consider a uniform electric field within the cavity. In order to compare the cases $N_{mol}=1$ and $N_{mol}=2$, we apply a scaling factor $1/\sqrt{N_{mol}}$ on the collective coupling strength $\bm{\lambda}_c$ to obtain a fixed Rabi splitting.
\begin{equation}
\label{eq:coupling}
\bm{\lambda}_{c} = \frac{\lambda_0}{\sqrt{N_{mol}}} \bm{e}_c
\end{equation} 
Here $\lambda_0$ is equivalent to $\lambda_c$ in Eq.~\eqref{eq:lam} in the single-molecule case.
As a result, we increase the mode volume $V_c$ of the cavity, but by including more molecules, we keep the average density of molecules $N_{mol}/V_c$ fixed. We use a coupling strength $\lambda_0$ of \SI{0.03}{\au}, which corresponds to a cavity electric field strength of \SI{1.5}{\volt\per\nano\meter} in a Fabry-P\'erot-like setup.

By scanning along the bond length of each \ce{HF} molecule $r_{HF}$ and the photon displacement coordinate $q_c$, we construct the $(N_{mol}+1)$-dimensional \gls{cpes} together with the corresponding dipole moment surfaces. 
The potential energy surfaces for the cases $N_{mol}=1$ and $N_{mol}=2$ are interpolated to an equally spaced grid of 128$\times$64 ($r_{HF}\times q_{c}$) grid points and 128$\times$128$\times$64 ($r_{HF}\times r_{HF} \times q_{c}$) grid points, respectively.
A Gaussian-shaped trial function is numerically propagated in imaginary time ~\cite{Kosloff1986-ow} (time step \SI{0.1}{\au} and 70000 time steps) on \gls{cpes} with the Arnoldi propagation scheme~\cite{Smyth1998-dv} to obtain the first ten nuclear-photonic eigenfunctions and to determine the corresponding transition dipole moments. All quantum dynamics simulations are performed with the open source quantum dynamics code QDng~\cite{Kowalewski2024-ue}.
All calculations were performed in a reproducible environment using the Nix package manager together with NixOS-QChem \cite{nix} (commit f5dad404) and Nixpkgs (nixpkgs, 22.11, commit 594ef126).

\section{Results}

In the following, we present \gls{dqc} spectra simulated for a single \ce{HF} molecule and a pair of \ce{HF} molecules resonantly coupled to a single-photon mode of an optical cavity. 
 Self-consistent treatment within the \gls{cbohf} ansatz allows the electronic structure
of the molecular ensemble to respond to the cavity field. Moreover, the inclusion of \gls{dse} terms allows for a proper description of the self-polarization of the coupled cavity-molecular system~\cite{Sidler2022-cg,Schnappinger2023-hh,Sidler2024-vm}. 

\subsection{Vibrational polaritons in hydrogen fluoride molecules}\label{subsec:polaritons}

\begin{figure}
     \centering
    \includegraphics[width=0.65\textwidth]{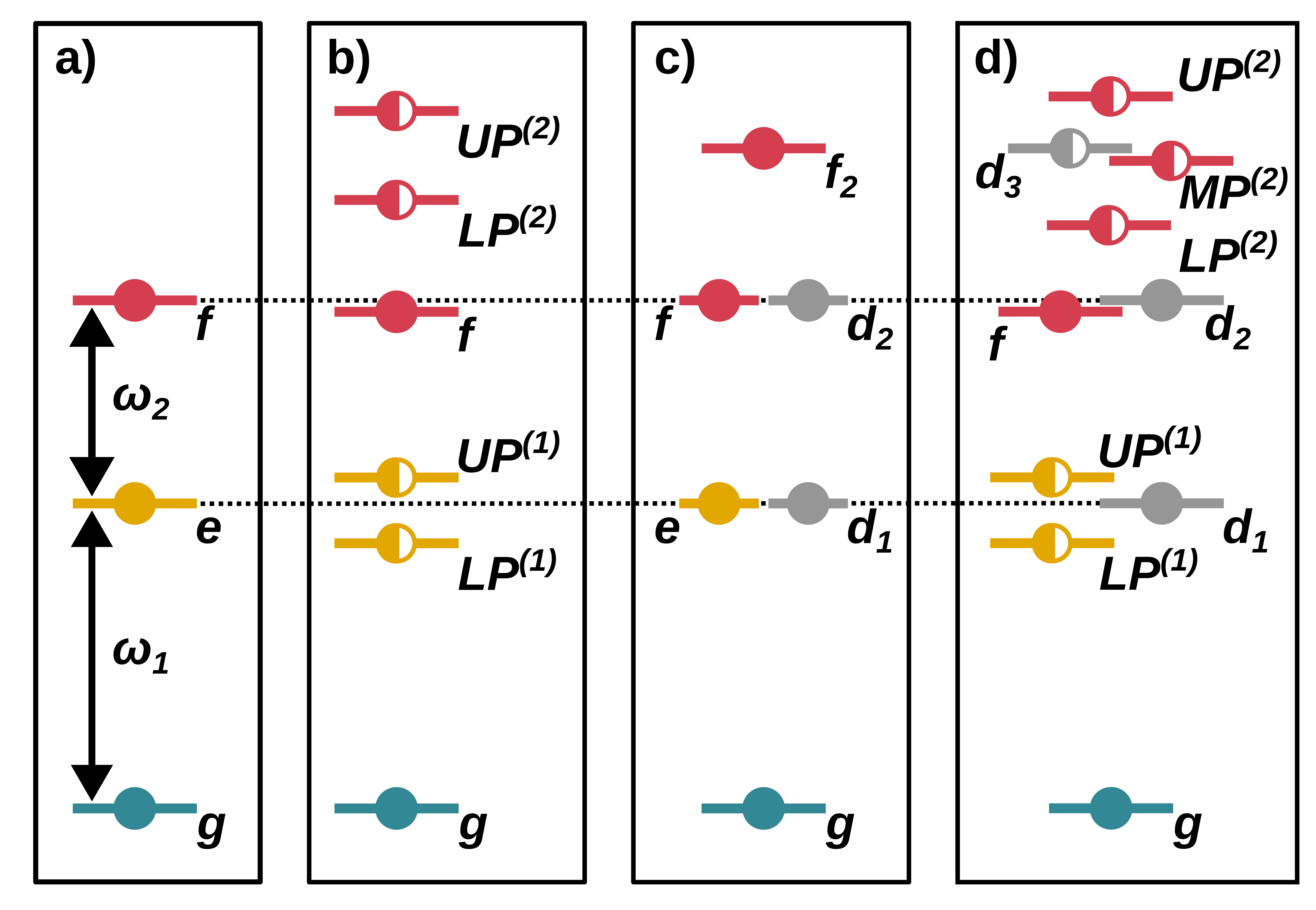}
    \caption{Schematic energy level diagrams for a single \ce{HF} molecule and a pair of \ce{HF} molecules without a cavity in a) and c) and resonantly coupled with a single cavity mode in b) and d). The cavity frequency $\omega_c$ is resonant with $\omega_1$. The ground state is colored green, the single excited state is colored yellow, and the double excited states are colored red. The optically dark states originating from the ground state to the single excited manifold and from the single excited manifold to the double excited manifold are shown in gray. Full circles indicate states of predominantly matter character and half-filled circles indicate states with a mixed matter and photon contribution.} 
 \label{fig:4281_levels}
\end{figure}

Figure~\ref{fig:4281_levels} shows schematic energy level diagrams for a single \ce{HF} molecule and two \ce{HF} molecules without a cavity in a) and c) and resonantly coupled with a single cavity mode ($\omega_c = \omega_1$) in b) and d) respectively. Apart from the expected energetic differences between the results obtained with the three different energy expectation values, the schematic energy-level diagrams shown are the same in all three cases.
An analysis of the discussed polaritonic states for both the single-molecule case and the two-molecule case in terms of uncoupled bare states $\ke{v,n}$ can be found in the Supporting Information section~S1. 
In the notation used for the bare states, $v$ describes the molecular vibrational excitation of the uncoupled system and $n$ is the uncoupled photon number.

In case of a single uncoupled \ce{HF} molecule, the energy diagram shown in Fig.~\ref{fig:4281_levels}~a), consists of the vibrational ground state $g$, a single excited state $e$ and a double excited state $f$, which correspond to vibrational quantum numbers $v=0, 1, 2$ respectively.
These states are separated by $\omega_{1} = \SI{4281}{\per\centi\meter}$ and $\omega_{2} = \SI{4108}{\per\centi\meter}$, respectively. The corresponding molecular anharmonicity $\Delta = \omega_{1} - \omega_{2}$ for \ce{HF} is \SI{173}{\per\centi\meter}. 
When a single-cavity photon mode strongly couples to a single \ce{HF} molecule aligned with the cavity polarization axis, two single excited states and three double excited states in the system are formed, see Fig.~\ref{fig:4281_levels}~b).
The cavity frequency $\omega_{c}$ is chosen to be resonant with the fundamental transition $\omega_{1} = \SI{4281}{\per\centi\meter}$ and the coupling strength $\lambda_c$ is \SI{0.03}{\au}. 
A complementary analysis for the case $\omega_{c} = \omega_{2}$, corresponding to the first hot transition, can be found in the Supporting Information section~S2.
Within the single excitation manifold, this resonant interaction leads to the expected formation of a pair of hybrid polaritonic states $UP^{(1)}$ and $LP^{(1)}$ with a Rabi splitting $\Omega_R^{(1)}$ of \SI{60}{\per\centi\meter} for the chosen coupling strength.
The lowest energy state $f$ in the second excitation manifold of the coupled cavity-molecule system is characterized mainly by a double excitation in the molecular part ($v=2$, bare state $\ke{2,0}$).
Compared to its cavity-free counterpart, this state is stabilized by approximately \SI{20}{\per\centi\meter}. 
The remaining two states are on average $\Delta$ higher in energy than $f$ and form a second pair of hybrid polaritonic states $UP^{(2)}$ and $LP^{(2)}$ with a Rabi splitting $\Omega_R^{(2)}$ of $\SI{90}{\per\centi\meter}$. 
As expected, the Rabi splitting $\Omega_R^{(2)}$ in the second excitation manifold is increasing by approximately a factor of $\sqrt{2}$ compared to $\Omega_R^{(1)}$.
$UP^{(2)}$ and $LP^{(2)}$ are formed by the bare state $\ke{1,1}$, in which the molecule and the cavity are both singly excited, and the bare state $\ke{0,2}$, in which the cavity is doubly excited.
None of the states in this coupled cavity-molecule system is expected to be dark with respect to the ground state and the single excited manifold. %\CF{The increase of the Rabi splitting from 60 to 90 is due to a factor $\sqrt{2}$ because we are coupled to the $\ke{0;2}$ state, right ? }

In the case of two \ce{HF} molecules, in order to describe the bare molecular states, we use the basis $\ke{v_s,v_a}$ formed by the two quantum numbers $v_s$ and $v_a$. The quantum numbers $v_s$ and $v_a$ correspond to the vibrational excitation of the symmetric and antisymmetric stretching normal modes respectively. For more details, see Section~S1 of the supporting information. 
The two \ce{HF} molecules are oriented parallel with respect to the cavity polarization axis and separated by \SI{800}{\angstrom} to ensure that the non-cavity-induced intermolecular couplings are negligible.
Figure~\ref{fig:4281_levels}~c) shows the corresponding energy level diagram. 
The two single excited states $e$ and $d_1$ are energetically degenerate.
Both correspond to the first excitation of the symmetric stretching mode ($\ke{10}$) and the antisymmetric stretching mode ($\ke{01}$) of the two \ce{HF} molecules, respectively. 
This explains why the $e$ state (symmetric mode) can be excited from the ground state, while the other state is dark because of symmetry. 
Similarly, the two double excited states $f$ and $d_2$ are energetically degenerate and correspond to the second excitation of  symmetric stretching mode and the antisymmetric stretching mode $\ke{20}$ and $\ke{02}$. Consequently, the $f$ state is a bright state while $d_2$ is dark.
The remaining state $f_2$ is $\Delta$ higher in energy and is formed by a simultaneous single excitation of both the symmetric and antisymmetric stretching mode $\ke{11}$.

By resonantly coupling the two \ce{HF} molecules to the single cavity mode, again a pair of hybrid polaritonic states $UP^{(1)}$ and $LP^{(1)}$ with a Rabi splitting $\Omega_R^{(1)}$ of \SI{60}{\per\centi\meter} is formed (see Fig.~\ref{fig:4281_levels}~d)).
Since the $d_1$ state (antisymmetric stretching mode $\ke{01}$) is dark, it is not affected by the coupling to the cavity.
The double excitation manifold is more complex, with a total of six states. 
Both $f$ ($\ke{20}$) and $d_2$ ($\ke{02}$) are only slightly or not affected by the cavity interaction and do not change their predominantly molecular character.
The four remaining states, $UP^{(2)}$, $MP^{(2)}$, $d_3$, and $LP^{(2)}$, all have a photonic contribution, see TABLE~S2 in Section~S1 of the Supporting Information.
However, only $UP^{(2)}$, $MP^{(2)}$, and $LP^{(2)}$ are true hybrid polaritonic states, while the state $d_3$ ($\ke{01,1}$) is formed by adding a cavity excitation to the dark state $d_1$  ($\ke{01,0}$) without further hybridization.
The splitting between the three polaritonic states is \SI{53}{\per\centi\meter}, corresponding to a total Rabi splitting $\Omega_R^{(2)}$ of \SI{106}{\per\centi\meter}.
The two states $UP^{(2)}$ and $LP^{(2)}$ are mainly characterized by a hybridization of the two bare states $\ke{10,1}$ and $\ke{00,2}$. The middle polariton $MP^{(2)}$ is mostly a linear combination of $\ke{00,2}$ and $\ke{11,0}$, which corresponds to $f_2$ in the case of two uncoupled molecules Fig.~\ref{fig:4281_levels}~c).

\subsection{\gls{dqc} spectra for a single \ce{HF} molecule}

\begin{figure}
     \centering
    \includegraphics[width=1.0\textwidth]{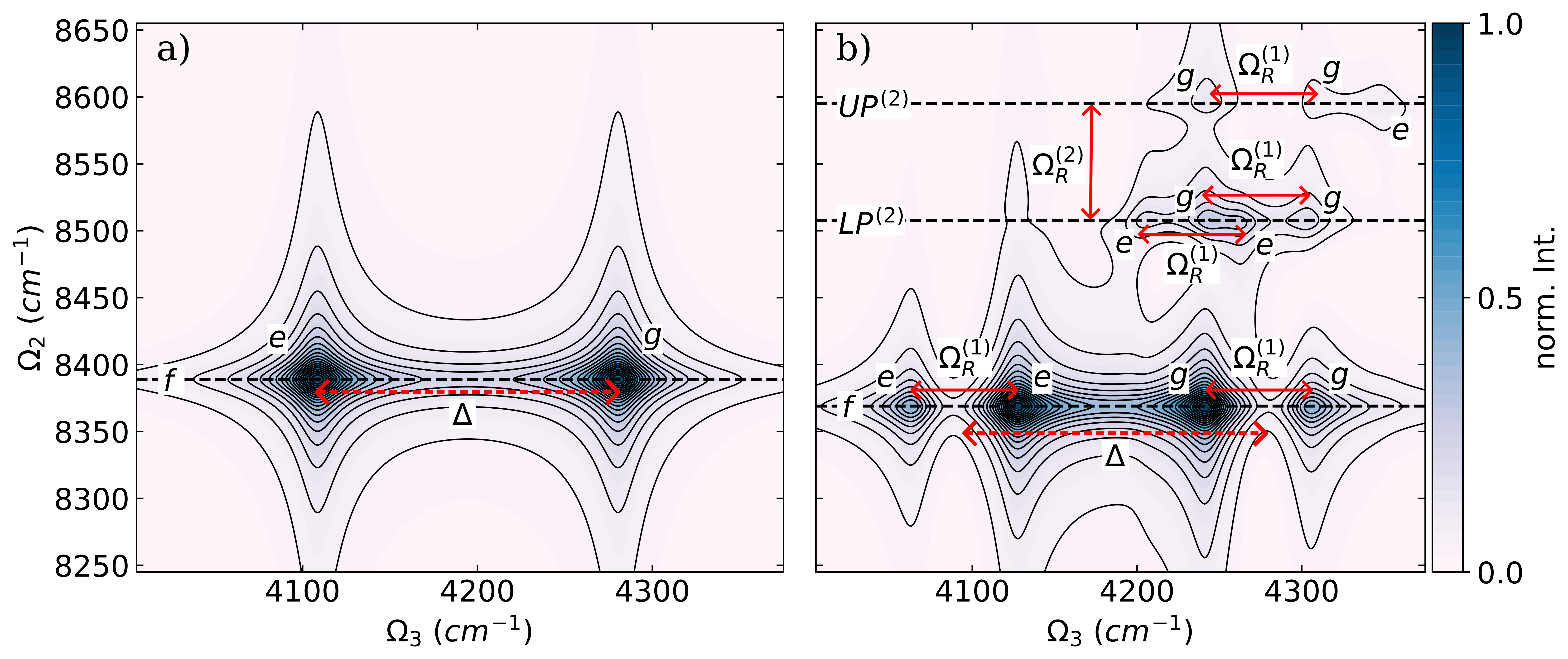}
    \caption{Absolute value of the normalized \gls{dqc} spectra of a single \ce{HF} molecule a) without a cavity and b) coupled to a single cavity mode with $\omega_{c} = \SI{4281}{\per\centi\meter} $. The coupling strength $\lambda_c$ is \SI{0.03}{\au} and the dephasing $\gamma$ is \SI{10}{\per\centi\meter}. The black horizontal dashed lines mark the energy of the final states, and all signals are labeled $e$ and $g$, indicating that the initial state is the ground state or an intermediate state. The red lines with arrows highlight relevant energy differences.} 
\label{fig:1hf_dqc}
\end{figure}

Figs.~\ref{fig:1hf_dqc}~a) and  ~\ref{fig:1hf_dqc}~b) show the absolute value of the \gls{dqc} spectra for a single \ce{HF} molecule without cavity and resonantly coupled with a single cavity mode, respectively.
The normalized absolute values are used in the following to simplify the comparison between spectra obtained with different \gls{cpes} and number of molecules. 

As illustrated in Fig.~\ref{fig:4281_levels}~a), the \ce{HF} molecule without cavity is a simple three-level system and the resulting \gls{dqc} spectra exhibit two peaks 
(see Fig.~\ref{fig:1hf_dqc}~a)).
Both peaks are resonant at \SI{8393}{\per\centi\meter} on the $\Omega_2$ axis and separated by the molecular anharmonicity $\Delta$ of \SI{173}{\per\centi\meter} along the $\Omega_3$ axis. 
The one at $\Omega_3=\SI{4281}{\per\centi\meter}$ is due to the $g \rightarrow e$ transition, and the signal at \SI{4108}{\per\centi\meter} corresponds to the $e\rightarrow f$ transition.
For the coupled single-molecule single-cavity mode system, we observe three distinct resonances (horizontal dashed lines) on the $\Omega_2$ axis in the \gls{dqc} spectra (see Fig.~\ref{fig:1hf_dqc}~b)).  
At $\Omega_2 = \SI{8369}{\per\centi\meter}$, corresponding to the final state $f$, we observe four peaks, which are also the most intense of the whole \gls{dqc} spectra. 
These peaks form a pair of doublets, where the two doublets are separated by the molecular anharmonicity $\Delta$ of \SI{173}{\per\centi\meter}. The two peaks of each doublet are separated by the Rabi splitting $\Omega_R^{(1)}$ of \SI{60}{\per\centi\meter}. 
The doublet around $\Omega_3 = \SI{4290}{\per\centi\meter}$ is due to the transition from the ground state to the pair of hybrid polaritonic states $UP^{(1)}$ and $LP^{(1)}$, and the other is due to the transition from these hybrid states to the $f$ state.
Comparing the \gls{dqc} spectra shown in Fig.~\ref{fig:1hf_dqc}~a) and~b) 
the stabilization of the state $f$ due to the cavity interaction is clearly visible. 
The other two resonances on the $\Omega_2$ axis associated with the two hybrid states $UP^{(2)}$ and $LP^{(2)}$ are weaker and separated by a Rabi splitting $\Omega_R^{(2)}$ of $\SI{90}{\per\centi\meter}$.
At $\Omega_2 = \SI{8507}{\per\centi\meter}$, final state $LP^{(2)}$, four peaks are distinguishable at \SI{4206}{\per\centi\meter} ($LP^{(1)} \rightarrow LP^{(2)}$), at \SI{4245}{\per\centi\meter} ($g\rightarrow LP^{(1)}$), at \SI{4260}{\per\centi\meter} ($UP^{(1)}\rightarrow LP^{(2)}$) and at \SI{4303}{\per\centi\meter} ($g\rightarrow UP^{(1)}$) for the chosen dephasing.
By careful analysis of the underlying transitions, these signals can be grouped into two doublets, with each peak approximately separated by the Rabi splitting $\Omega_R^{(1)}$ of \SI{60}{\per\centi\meter}. 
For the final state $UP^{(2)}$, $\Omega_2 = \SI{8595}{\per\centi\meter}$, only three very weak signals can be observed at \SI{4242}{\per\centi\meter} ($g\rightarrow LP^{(1)}$), at \SI{4305}{\per\centi\meter} ($g\rightarrow UP^{(1)}$) and at \SI{4348}{\per\centi\meter} ($LP^{(1)}\rightarrow UP^{(2)}$). The intensity of the fourth possible transition ($UP^{(1)}\rightarrow UP^{(2)}$) is too weak to be seen in the spectra. 
The first two peaks are again approximately split by the Rabi splitting $\Omega_R^{(1)}$.

\begin{figure}
     \centering
    \includegraphics[width=1.0\textwidth]{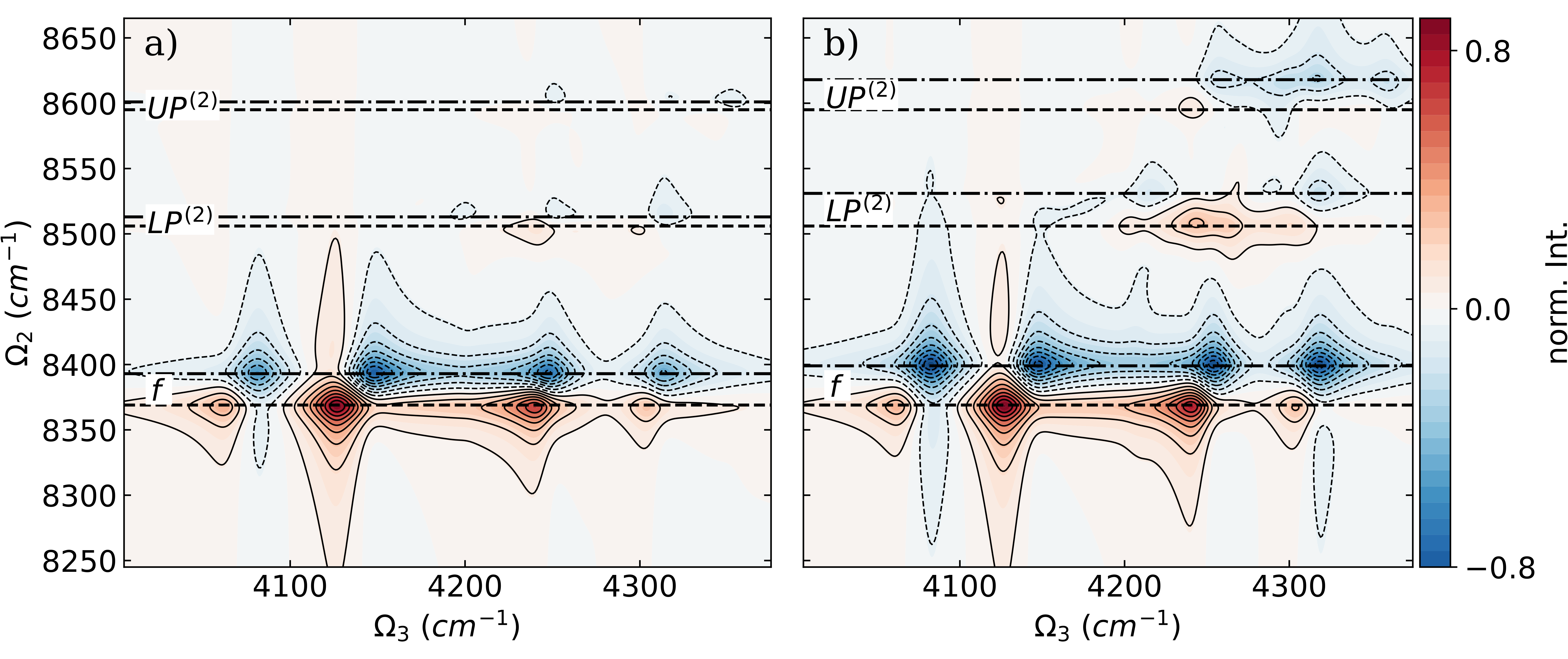}
    \caption{Difference of the \gls{dqc} signal $\Delta \mathcal S $ of a single \ce{HF} molecule coupled to a photon mode with $\omega_{c} = \SI{4281}{\per\centi\meter}$ between a)  full \gls{cbohf} and linear \gls{cbohf}  and b) full \gls{cbohf} and \gls{etc}. Individual \gls{dqc} spectra are normalized, and the absolute value is used to calculate the difference. The coupling strength $\lambda_c$ is \SI{0.03}{\au} for both frequencies and the dephasing $\gamma$ is \SI{10}{\per\centi\meter}. The energies of the final states are marked with black dashed lines for the full \gls{cbohf} case and with dashed dotted lines for the linear \gls{cbohf} and \gls{etc} cases, respectively.} 
\label{fig:1hf_dqc_delta}
\end{figure}

Comparisons between \gls{dqc} signals based on the full \gls{cbohf} surface, the linear \gls{cbohf} surface, and the \gls{etc} surfaces are shown as difference spectra $\Delta \mathcal S $ in Fig.~\ref{fig:1hf_dqc_delta}. $\Delta \mathcal S $ is calculated according to Eq.~\ref{eq:DeltaS1}
\begin{equation}
 \label{eq:DeltaS1}   
 \Delta \mathcal S = \left| \mathcal S_{CBO} \right|  - \left| \mathcal S_{CBO}^{lin} \right| \quad  \text{ and} \quad  \Delta \mathcal S = \left| \mathcal S_{CBO} \right|  - \left| \mathcal S_{ETC} \right| 
\end{equation}
In both difference spectra the peaks corresponding to the final state $f$ are the most affected. 
All peaks associated with the $f$ state are red-shifted by approximately \qtyrange{20}{30} {\per\centi\meter} in $\Omega_2$ and approximately \qtyrange{10}{20}{\per\centi\meter} in $\Omega_3$ when the \gls{dse} contribution is included, see Fig.~\ref{fig:1hf_dqc_delta}~a), or the \gls{scf} procedure is performed, see Fig.~\ref{fig:1hf_dqc_delta}~b)).  
As shown in Fig.~\ref{fig:1hf_dqc_delta}~a) the $UP^{(2)}$ resonance and the $LP^{(2)}$ resonance and the corresponding peaks are almost unaffected when the \gls{dse} contribution is not included.
In contrast, when the \gls{etc} surface is used to determine the \gls{dqc} spectra, the resonances of $UP^{(2)}$ and the resonances of $LP^{(2)}$ are affected, see Fig.~\ref{fig:1hf_dqc_delta}~b). Both are blue shifted by approximately \SI{20}{\per\centi\meter} in $\Omega_2$ for the case without \gls{scf}. Also, the intensity distribution changes, $UP^{(2)}$ loses intensity while $LP^{(2)}$ gains it when going from the \gls{etc} spectra to the full \gls{cbohf} results. 

For completeness, the real ($Re$) and imaginary ($Im$) parts of the \gls{dqc} spectra of the coupled single-molecule single-cavity mode system are shown in the supporting information section S3, Fig.~S6.

\subsection{\gls{dqc} spectra for a pair of \ce{HF} molecules}

\begin{figure}
     \centering
\includegraphics[width=1.0\textwidth]{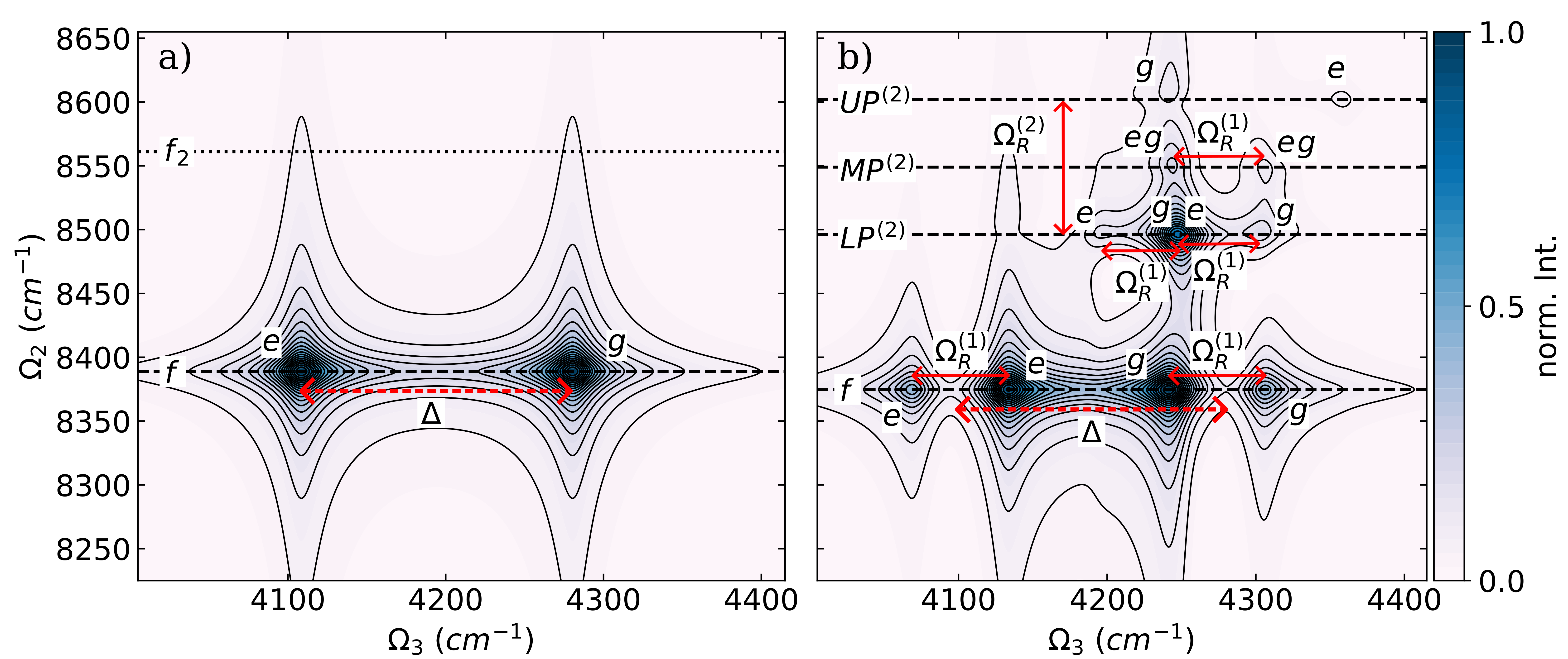}
    \caption{Absolute value of the normalized \gls{dqc} spectra of two parallel oriented \ce{HF} molecules a) without a cavity and b) coupled to a single cavity mode with $\omega_{c} = \SI{4281}{\per\centi\meter} $. The coupling strength $\lambda_0$ is \SI{0.03}{\au} and the dephasing $\gamma$ is \SI{10}{\per\centi\meter}. The black horizontal dashed lines mark the energy of the final states, and all signals are labeled $e$ and $g$, indicating that the initial state is the ground state or an intermediate state. The red lines with arrows highlight relevant energy differences.} 
\label{fig:2hf_dqc}
\end{figure}

The absolute value of the normalized \gls{dqc} spectra for two identical \ce{HF} molecules are shown in Fig.~\ref{fig:2hf_dqc} for the case without a cavity in a) and resonantly coupled with a single cavity mode in b).

The \gls{dqc} spectra of two parallel oriented \ce{HF} molecules not interacting with a cavity field is shown in Fig.~\ref{fig:2hf_dqc}~a) and is nearly identical to the single-molecule situation, see Fig.~\ref{fig:1hf_dqc}~a) for comparison.
The bright state $f_2$, which is described by a simultaneous single excitation of both the symmetric and antisymmetric stretching mode, is not visible because of an exact cancellation of the two Liouville paths. The corresponding energy splittings between the $g \rightarrow e$ transition and the $e \rightarrow f_2$ transition are identical, leading to a harmonic ladder that is not visible in a \gls{dqc} spectrum. 
A black dotted line in Fig.~\ref{fig:2hf_dqc}~a) indicates the energetic position of the corresponding hypothetical $f_2$ resonances.
The two additional vibrational states ($d_1$ and $d_2$) present in the case of two molecules are dark, as illustrated in Fig.~\ref{fig:4281_levels}~c), since they are associated with excitations of the antisymmetric linear combination of the two stretching modes. 
As a consequence, the resulting \gls{dqc} spectrum has only two peaks and, thus representing an effective three-level system. 
However, once the two \ce{HF} molecules are resonantly coupled to the cavity mode, the resulting \gls{dqc} spectrum, shown in Fig.~\ref{fig:2hf_dqc}~b), is clearly distinguishable from the single molecule case. 
We observe four distinct resonances (horizontal dashed lines) on the $\Omega_2$ axis in the \gls{dqc} spectra, since the other two possible double excited states $d_2$ and $d_3$ are dark for symmetry reasons.  
Similarly to the case of a single \ce{HF} molecule, we observe four peaks at $\Omega_2 = \SI{8374}{\per\centi\meter}$, corresponding to the final state $f$, which are also the most intense of the whole \gls{dqc} spectrum. 
Interestingly, the resonance of the $f$ state is nearly identical in both the one-molecule and two-molecule cases. 
The four associated peaks can again be grouped as a pair of doublets separated by the molecular anharmonicity $\Delta$ of about \SI{173}{\per\centi\meter} and the Rabi splitting $\Omega_R^{(1)}$ of \SI{65}{\per\centi\meter}. 
Due to the rescaling of the coupling strength according to Eq.~\ref{eq:coupling} $\Omega_R^{(1)}$ is similar to the one-molecule case.
The doublet around $\Omega_3 = \SI{4274}{\per\centi\meter}$ is due to the transition from the ground state to the pair of hybrid polaritonic states $UP^{(1)}$ and $LP^{(1)}$, and the other one is due to the transition from these hybrid states to the $f$ state.
The three hybrid polaritonic states $UP^{(2)}$, $MP^{(2)}$, and $LP^{(2)}$ have visible resonances on the $\Omega_2$ axis.
These three resonances are separated by about \SI{53}{\per\centi\meter}, corresponding to a total Rabi splitting $\Omega_R^{(2)}$ of \SI{106}{\per\centi\meter}, consistent with the observed increase of the Rabi splitting in the single molecule case. 
At $\Omega_2 = \SI{8496}{\per\centi\meter}$, final state $LP^{(2)}$, we see an intense peak at \SI{4248}{\per\centi\meter} comparable to the peaks corresponding to the final state $f$ and two weaker signals at \SI{4197}{\per\centi\meter} ($LP^{(1)} \rightarrow LP^{(2)}$) and at \SI{4303}{\per\centi\meter} ($g \rightarrow UP^{(1)}$). 
These three signals are approximately separated by the Rabi splitting $\Omega_R^{(1)}$.
The high intensity of the central peak can be explained by the overlap of the $g \rightarrow LP^{(1)}$ and $UP^{(1)} \rightarrow LP^{(2)}$ transitions. 
For the final state $MP^{(2)}$, $\Omega_2 = \SI{8551}{\per\centi\meter}$, only two signals can be observed at \SI{4244}{\per\centi\meter} and at \SI{4306}{\per\centi\meter} separated again by the Rabi splitting $\Omega_R^{(1)}$. 
These two signals are formed by the transitions $g \rightarrow LP^{(1)}$ and $LP^{(1)} \rightarrow MP^{(2)}$ and the transitions $UP^{(1)} \rightarrow MP^{(2)}$ and $g \rightarrow UP^{(1)}$ respectively. 
As mentioned in section~\ref{subsec:polaritons} the final state $MP^{(2)}$ is formed by hybridization of the two bare states $\ke{00,2}$ and $\ke{11,0}$. 
Without coupling to a cavity, this $\ke{11,0}$ state, labeled $f_2$ in Fig.~\ref{fig:2hf_dqc}~a), is not visible because of a cancellation of the Liouville diagrams. In the cavity, however, this state becomes visible due to the hybridization, which induces anharmonicity in the corresponding  excitation ladder. 
The identical process is observed when the cavity is resonant with the first hot transition and discussed in Section~S2 of the supporting information. 
Similarly to the single-molecule case, only two very weak signals corresponding to the final state $UP^{(2)}$, $\Omega_2 = \SI{8610}{\per\centi\meter}$, can be observed at \SI{4242}{\per\centi\meter} ($g\rightarrow LP^{(1)}$) and at \SI{4359}{\per\centi\meter} ($UP^{(1)}\rightarrow UP^{(2)}$). All other possible transitions are too weak to be observed in the spectra.
There is no clear indication that any of the three dark states $d_1$ and $d_2$ takes part in the signal and thus these states are not needed to explain the \gls{dqc} signals.
As for the single molecule case, for completeness we show the real ($Re$) and imaginary ($Im$) parts of the \gls{dqc} spectra of the two-molecule case coupled to a single-cavity mode in the supporting information section S3 Fig.~S7.

\begin{figure}
     \centering
    \includegraphics[width=1.0\textwidth]{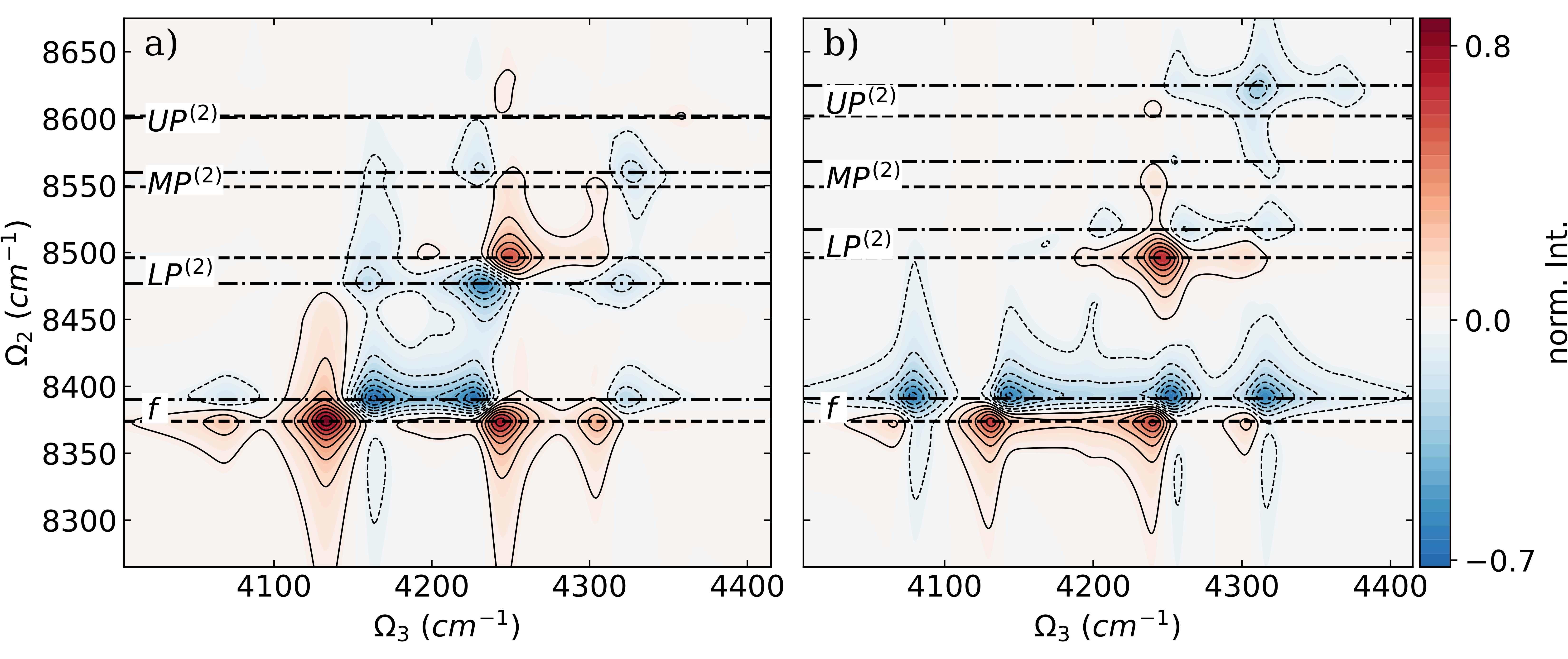}
    \caption{Difference $\Delta \mathcal S $ of the \gls{dqc} signal of two parallel \ce{HF} molecule coupled to a photon mode with $\omega_{c} = \SI{4281}{\per\centi\meter}$ between a)  full \gls{cbohf} and linear \gls{cbohf}  and b) full \gls{cbohf} and \gls{etc}. Individual \gls{dqc} spectra are normalized, and the absolute value is used to calculate the difference. The coupling strength $\lambda_c$ is \SI{0.03}{\au} for both frequencies and the dephasing $\gamma$ is \SI{10}{\per\centi\meter}.  The energies of the final states are marked with black dashed lines for the full \gls{cbohf} case and with dashed dotted lines for the linear \gls{cbohf} and \gls{etc} cases, respectively.} 
\label{fig:2hf_dqc_delta}
\end{figure}

In Fig.~\ref{fig:2hf_dqc_delta} the comparison between the \gls{dqc} signals based on the full \gls{cbohf} surface, the linear \gls{cbohf} surface, and the \gls{etc} surfaces for the two-molecule case is shown as difference spectra $\Delta \mathcal S $ calculated according to Eq.~\ref{eq:DeltaS1}. 
In agreement with the single-molecule case, the peaks corresponding to the state $f$ are most affected in both difference spectra.
For the comparison between the full \gls{cbohf} and linear \gls{cbohf} results shown in Fig.~\ref{fig:2hf_dqc_delta}~a), the $f$ peaks are only red-shifted by about \SI{20}{\per\centi\meter} in $\Omega_2$, while no general trend for shifts in $\Omega_3$ is apparent. 
When the \gls{dse} term is not included (see Fig.~\ref{fig:2hf_dqc_delta}~a)), the resonances of the two hybrid polaritonic states $MP^{(2)}$ and $LP^{(2)}$ are affected. 
In particular, for the $LP^{(2)}$ state, a red shift is observed in both $\Omega_2$ and $\Omega_3$ for the strong signal around \SI{4248}{\per\centi\meter} without \gls{dse} included. 
Interestingly, the weak $UP^{(2)}$ remains almost unchanged, similar to the case of a single molecule. 
In contrast, when the \gls{etc} surface is used to determine the \gls{dqc} spectra, the observed changes in the difference spectra in the single-molecule case and the two-molecule case are quite similar. 
For the $f$ signals, almost the same red shift is observed in both $\Omega_2$ and $\Omega_3$, compare Fig.~\ref{fig:1hf_dqc_delta}~b) and Fig.~\ref{fig:2hf_dqc_delta}~b).
The resonances of three hybrid polaritonic states $UP^{(2)}$, $MP^{(2)}$, and $LP^{(2)}$ are all blue shifted by roughly \SI{20}{\per\centi\meter} in $\Omega_2$ for the \gls{etc} case. 
Consistent with the single-molecule results, the intensity distribution changes, $UP^{(2)}$ loses intensity while $LP^{(2)}$ gains it when going from the \gls{etc} spectra to the full \gls{cbohf} results. 
In general, the observed effect on the \gls{dqc} spectra of the \gls{scf} treatment is the same for one and two \ce{HF} molecules, where the \gls{dse} included in the \gls{cbohf} approach shows additional changes in the spectra when going from one to two molecules. 

\begin{figure}
     \centering
    \includegraphics[width=1.0\textwidth]{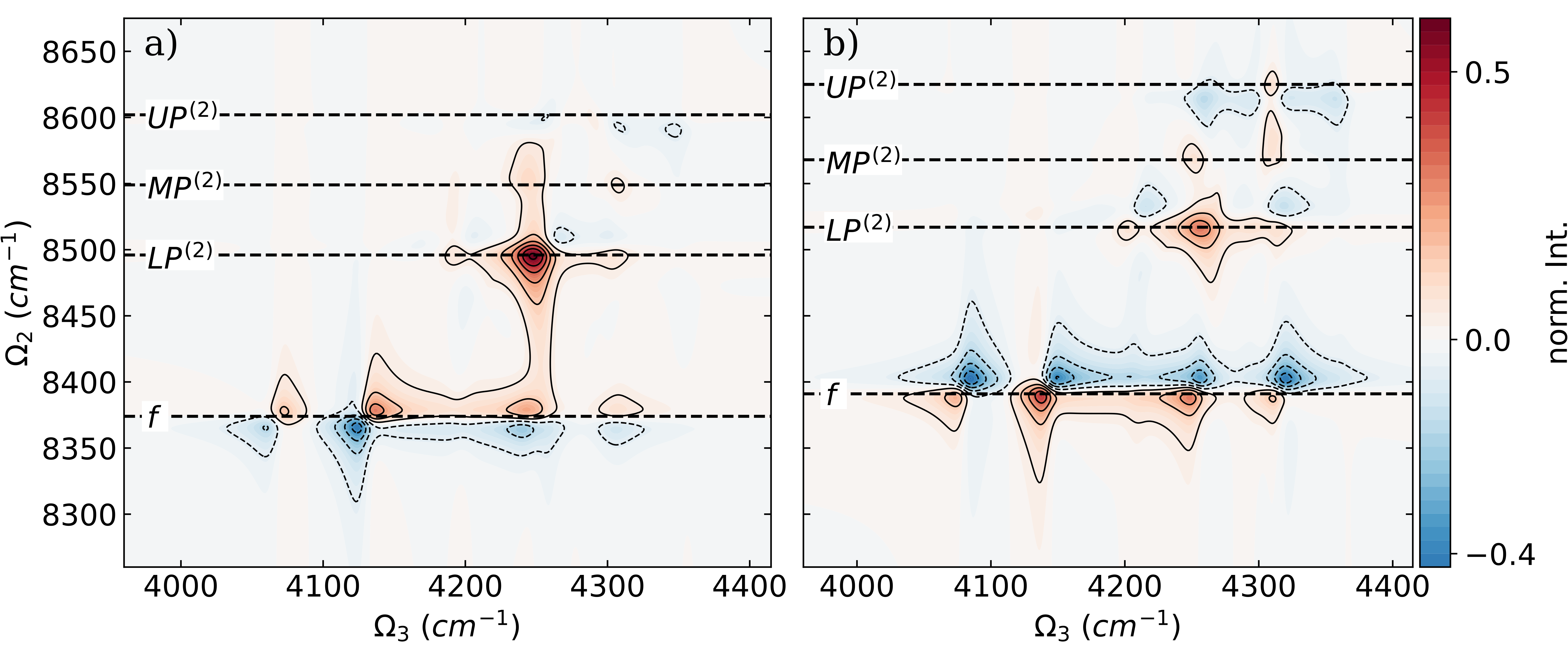}
    \caption{Difference of the absolute values of the \gls{dqc} spectra of two parallel oriented \ce{HF} molecules and a single \ce{HF} molecule coupled to a photon mode with $\omega_{0} = \SI{4281}{\per\centi\meter}$, 
    $\lambda_c = \SI{0.03}{\au}$ and the dephasing $\gamma$ is \SI{10}{\per\centi\meter}. a) The full \gls{cbohf} \glspl{cpes} are used to construct the \gls{dqc} spectra, and b) the \gls{etc} model \glspl{cpes} are used. The energies of the final states are marked with black dashed lines for the two molecule case.} 
\label{fig:2hf_dqc_2p_1p}
\end{figure}

As a final step, we want to further analyze the differences in the \gls{dqc} spectra when going from one to two \ce{HF} molecules for the full \gls{cbohf} approach and using the \gls{etc} model. 
The corresponding difference spectra are shown in Fig.~\ref{fig:2hf_dqc_2p_1p}
Analyzing the difference spectrum in Fig.~\ref{fig:2hf_dqc_2p_1p}~a) for the full \gls{cbohf} case, two main changes appear when going from one to two \ce{HF} molecules. 
The most striking change is the increase in intensity for the signal around \SI{4248}{\per\centi\meter}, which corresponds to the final state $LP^{(2)}$.
As discussed for the two-molecule \gls{dqc} spectrum shown in Fig.~\ref{fig:2hf_dqc}~b), this intense peak is due to two overlapping signals that are still separated in the single-molecule \gls{dqc} spectrum (see Fig.~\ref{fig:1hf_dqc}~b)).
The other change is a blue shift in $\Omega_2$ of all signals corresponding to the $f$ resonance. 
For the \gls{etc} case shown in Fig.~\ref{fig:2hf_dqc_2p_1p}~b) all peaks associated with the final state $f$ are shifted to lower frequencies in both $\Omega_2$ and $\Omega_3$. 
With respect to polaritonic sates, only smaller changes are observed when going from one to two \ce{HF} molecules. 
The distinct differences in the comparison between the full \gls{cbohf} results and those obtained using the \gls{etc} model are due solely to the \gls{scf} process, which allows the electronic structure to respond to the cavity field mode. 
This response seems to be particularly relevant when going beyond a single-molecule situation because of cavity-induced molecular interactions. 

\section{Summary and Conclusion}

Based on the recently formulated cavity Born-Oppenheimer Hartree-Fock ansatz~\cite{Schnappinger2023-hh,Schnappinger2023-wp} and inspired by the work of Saurabh and Mukamel~\cite{Saurabh2016-tc} we simulated two-dimensional \gls{dqc} spectra for the rather anharmonic case of one and two diatomic hydrogen fluoride \ce{HF} molecules coupled to an infrared cavity.
In both cases, the molecular system is coupled to a single-photon mode that is resonant with the first vibrational transition in \ce{HF}. 
Using ab initio \gls{cpes} at the \gls{cbohf} level of theory, we could demonstrate how ground-state vibrational excitation manifolds are modified upon coupling to a cavity mode. 
As a consequence, the molecular anharmonicities, the molecule-cavity interaction, and the self-polarization of the molecule-cavity system are naturally included.
Even in the rather simple diatomic case of \ce{HF}, coupling to an optical cavity results in a rather complex and "asymmetric" \gls{dqc} signal compared to model systems reported in the literature~\cite{Saurabh2016-tc}.
This complexity comes from the non-idealized second excitation manifold, where due to the anharmonicity of \ce{HF} both nearly pure molecular states and the expected hybrid polaritonic states contribute, as shown in Fig.~\ref{fig:4281_levels}.
The significantly different character of these states leads to a mixture of rather weak and rather strong transitions. 
We were able to show that neglecting the \gls{dse} contribution or not performing the \gls{scf} procedure has a significant effect on the \gls{dqc} signals. 
In both cases, the resonances of the $f$ state are affected, which is mainly an excited molecular final state. 
Therefore, it allows us to monitor the effect of \gls{dse} and \gls{scf} on the \gls{lp} and \gls{up} states in the single excitation manifold. 
However, for the single-molecule case without \gls{scf}, the $UP^{(2)}$ state and the $LP^{(2)}$ state also change.
For the case of two \ce{HF} molecules, the \gls{dse} contribution not only influences the $f$ resonances but also leads to a strong signal corresponding to the final state $LP^{(2)}$, which is clearly weaker without \gls{dse} or \gls{scf}.
This $LP^{(2)}$ signal is also the striking difference when going from a single molecule to a pair of molecules. In contrast, the resonances of the $f$ state are nearly identical as the number of molecules increases. 
An important result of our study is that we show that the \gls{dqc} spectrum is highly sensitive to the effect of the \gls{dse}. In particular, when considering the case of two molecules without a cavity, the  molecular state $\ke{11}$ corresponding to two simultaneously excited molecules do not contribute to the \gls{dqc} spectrum due to the fact that the two molecules do not interact. However, when considering two molecules inside the cavity, the cavity induce an effective interaction between the two molecules resulting from the \gls{dse} which can be directly probed by the \gls{dqc} spectrum through the MP$^{(2)}$ state
Therefore, we show that the \gls{dqc} technique offer a unique insight into the 
cavity-mediated intermolecular interactions. \gls{dqc} spectroscopy can be used to have a better understanding of many-body effects in molecular systems under vibrational strong coupling which in turn may thus provide a deeper mechanistic insight into chemical reactions under vibrational strong coupling.

\begin{acknowledgments}
This project has received funding from the European Research Council (ERC) under the European Union’s Horizon 2020 research and innovation program (grant agreement no. 852286).
Support from the Swedish Research Council (Grant No. VR 2022-05005) is acknowledged.
\end{acknowledgments}

\section*{Supplementary Material}
See the supplementary material for an analysis of the vibrational polaritonic eigenstates in terms of bare states, \gls{dqc} spectra for a cavity resonant with the first hot transition of the \ce{HF} molecule and the real and imaginary parts of the \gls{dqc} spectra.

\section*{Author Declaration Section}
\subsection*{Conflict of Interest Statement }
The authors have no conflicts to disclose

\subsection*{Author Contributions}
\textbf{Thomas Schnappinger}: Data curation (lead); Formal analysis (lead); Conceptualization (equal);  Investigation (equal); Methodology (equal); Visualization (lead); Writing – original draft (lead); Writing – review \& editing (equal). \textbf{Cyril Falvo}: Conceptualization (equal); Investigation (equal); Writing – review \& editing (equal). \textbf{Markus Kowalewski}: Conceptualization (equal); Formal analysis (equal); Funding
acquisition (lead); Methodology (equal);
Project administration (lead); Supervision (lead); Writing – original draft (equal); Writing – review \& editing (equal).

\section*{Data Availability}
The data that support the findings of this study are available from the corresponding author upon reasonable request.

\bibliography{lit.bib}

\end{document}

\title{Supporting Information: \\Disentangling collective coupling in vibrational polaritons  with double quantum coherence spectroscopy}

\author{Thomas Schnappinger}
\email{thomas.schnappinger@fysik.su.se}
\affiliation{Department of Physics, Stockholm University, AlbaNova University Center, SE-106 91 Stockholm, Sweden}
\author{Cyril Falvo}
\affiliation{Universit\'e Paris-Saclay, CNRS, Institut des Sciences Mol\'eculaires d’Orsay, 91405 Orsay, France}
\affiliation{Universit\'e Grenoble-Alpes, CNRS, LIPhy, 38000 Grenoble, France}
\author{Markus Kowalewski}
\email{markus.kowalewski@fysik.su.se}
\affiliation{Department of Physics, Stockholm University, AlbaNova University Center, SE-106 91 Stockholm, Sweden}

\date{\today}%

\maketitle

\tableofcontents
\clearpage

\section{Analysis of the vibrational polaritonic eigenstates}

\begin{figure} [ht]
\centering
    \includegraphics[width=0.5\textwidth]{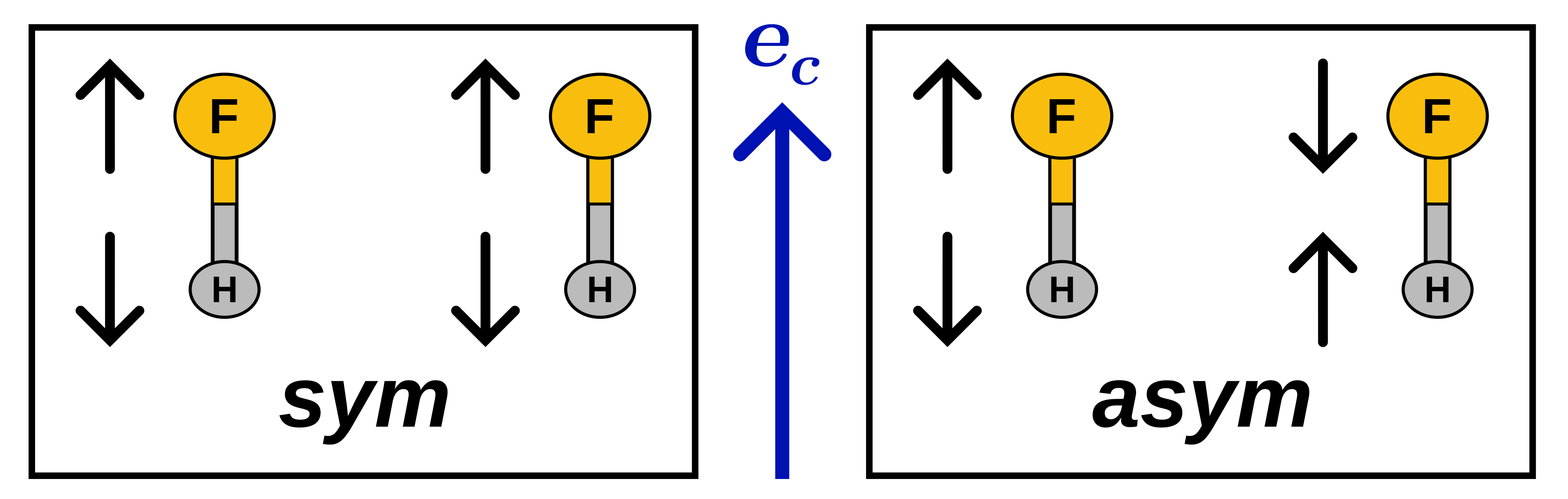}
    \caption{Sketch of the symmetric (sym) and antiymmetric (asym) liner combinations for two \ce{HF} molecules and the cavity polarization axis is shown in blue.} 
\label{fig:normel_modes}
\end{figure}

To analyze the polaritonic eigenstates for a single \ce{HF} molecule and a two \ce{HF} molecules coupled to a single-photon mode of an optical cavity, we extended the corresponding coupled eigenfunctions $\chi_{j}$ in terms of the uncoupled bare states:
\begin{equation}
\ke{\chi_{j}}  = \sum_i c_{i,j} \ke{v,n}_i
\end{equation}
where $v$ describes the vibrational excitation of the uncoupled molecular system and $n$ is the photon number.
The molecular part $v$ of the uncoupled molecular system is qualitatively described in terms of normal modes. In the case of two \ce{HF} molecules, the two included normal modes sketched in Fig.~\ref{fig:normel_modes} of the ensemble are the symmetric linear combination and the antiymmetric linear combination of the individual molecular stretching modes. We formulate $\ke{v}$ for two \ce{HF} molecules as $ \ke{v_{s} v_{a}}$, where $v_{s}$ and $v_a$ are the excitation numbers in the symmetric and antisymmetric stretching modes, respectively. 
Consequently, two molecular ensemble states $\ke{10}$ and $\ke{01}$ exist in the first excitation manifold, representing the first excitation of the symmetric stretching mode and the antisymmetric stretching mode, respectively.  The two states of the molecular ensemble $\ke{20}$ and $\ke{02}$ describe the corresponding second excitation in these stretching modes. The remaining double excited state $\ke{11}$ describes the simultaneous single excitation of the symmetric stretching mode and the antisymmetric stretching mode.
Apart from the numerical differences between the results obtained with the three different energy expectation values, the character of the polaritonic states expressed in terms of the bare states is qualitatively the same in all three cases.
Therefore, we will only show the results based on the full \gls{cbohf} energy expectation values. 
As a side note, we want to clarify that this basis is not complete but sufficient to qualitatively describe the light-matter hybridization in the first and second excitation manifold.

The absolute squares of the coefficients $|c_{i,j}|^2$ of all polaritonic states discussed in the case of a single \ce{HF} molecule coupled to a cavity with $\omega_{c} = \omega_{1} = \SI{4281}{\per\centi\meter}$ and a coupling strength $\lambda_c$ of \SI{0.03}{\au} are given in TABLE~\ref{tab:c2_hf1}.
The absolute squares of the coefficients $|c_{i,j}|^2$ of all polaritonic states discussed for the case of two \ce{HF} molecules coupled to a cavity with$\omega_{c} = \omega_{1} = \SI{4281}{\per\centi\meter}$ and a coupling strength $\lambda_c$ of \SI{0.03}{\au} are given in are given in TABLE~\ref{tab:c2_hf2}.
The corresponding coefficients $|c_{i,j}|^2$ of all polaritonic states discussed in section~S2 for the case of one and two \ce{HF} molecule and a cavity frequency $\omega_{c} = \omega_{2} = \SI{4108}{\per\centi\meter}$ are given in TABLE~\ref{tab:c2_hf1_w2} and TABLE~\ref{tab:c2_hf2_w2}.
\begin{table}[h]
    \centering
    \begin{tabular}{c|c c c c c c}
        & $\ke{0,0}$ & $\ke{1,0}$ & $\ke{0,1}$ & $\ke{2,0}$  & $\ke{0,2}$ & $\ke{1,1}$ \\
        \hline
        $\ke{g}$  &   \textbf{0.9945}   &  0.0055  & 0.0000  & 0.0000   &  0.0000     & 0.0000  \\
        \hline           
        $\ke{LP^{(1)}}$  &  0.0030  & \textbf{0.5301} & \textbf{0.4582} & 0.0000  & 0.0059  & 0.0027    \\
        $\ke{UP^{(1)}}$  & 0.0025  & \textbf{0.4535}  & \textbf{0.5359}   & 0.0000     &  0.0050   & 0.0031   \\
        \hline           
        $\ke{f}$  & 0.0000  &  0.0000   & 0.0002  & \textbf{0.9637} & 0.0008   & 0.0287 \\  $\ke{LP^{(2)}}$  & 0.0000  & 0.0063 & 0.0024 & 0.0179  & \textbf{0.5502}  & \textbf{0.4088} \\
        $\ke{UP^{(2)}}$  & 0.0000  & 0.0046 & 0.0032 & 0.0121 & \textbf{0.4216}   & \textbf{0.5448} \\       
        \hline        
        \hline
    \end{tabular}
    \caption{The absolute squares of the coefficients $|c_{i,j}|^2$ of all polaritonic states for the case of a single resonantly coupled  \ce{HF} molecule. The coefficients of the main contributions are highlighted in bold. The values are obtained using the full \gls{cbohf} energy expectation values, a cavity frequency $\omega_{c} = \omega_{1} = \SI{4281}{\per\centi\meter}$ and a coupling strength $\lambda_c$ of \SI{0.03}{\au}.}
    \label{tab:c2_hf1}
\end{table}

\begin{table}[h]
    \centering
    \begin{tabular}{c|c c c c c c c c c c}
        & $\ke{00,0}$ & $\ke{01,0}$  &  $\ke{10,0}$ & $\ke{00,1}$ & $\ke{02,0}$  & $\ke{20,0}$  & $\ke{01,1}$ & $\ke{10,1}$ & $\ke{11,0}$ & $\ke{00,2}$\\
        \hline
        $\ke{g}$  & \textbf{0.9890} &  0.0000 & 0.0000 & 0.0109 & 0.0000 & 0.0000 & 0.0000 & 0.0000 & 0.0000 & 0.0000 \\
         \hline     

        $\ke{d_1}$  & 0.0000 & \textbf{0.9886} & 0.0000 & 0.0000 & 0.0000 & 0.0000 & 0.0111 & 0.0002 & 0.0000 & 0.0000 \\
         
        $\ke{LP^{(1)}}$  & 0.0060 & 0.0000 & \textbf{0.4544} & \textbf{0.5224} & 0.0000 & 0.0000 & 0.0000 & 0.0052 &  0.0000 & 0.0117 \\

        $\ke{UP^{(1)}}$  & 0.0049 & 0.0000 & \textbf{0.5341} & \textbf{0.4449} & 0.0000 & 0.0000 & 0.0000  & 0.0059 & 0.0000 & 0.0099 \\
        \hline   

        $\ke{d_2}$  & 0.0000 & 0.0001 &  0.0000 & 0.0000 & \textbf{0.9740} & 0.0000 & 0.0000  & 0.0138 & 0.0002 & 0.0000 \\  
        
        $\ke{f}$  & 0.0000 & 0.0000 &  0.0002 & 0.0000 & 0.0000 &  \textbf{0.9720} & 0.0002  & 0.0002  & 0.0150 & 0.0004 \\

        $\ke{d_3}$  & 0.0000 &  0.0111 &  0.0000 & 0.0000 &  0.0144 & 0.0000 & \textbf{0.9374} & 0.0147 & 0.0000 & 0.0000 \\     

        $\ke{LP^{(2)}}$  & 0.0000 & 0.0000 & 0.0053 & 0.0087 &  0.0000 & 0.0109 & 0.0069 & \textbf{0.4396} & \textbf{0.1326} & \textbf{0.3710} \\

        $\ke{MP^{(2)}}$ & 0.0000 & 0.0000  & 0.0001 & 0.0071 & 0.0000 & 0.0001 & 0.0001 & 0.0071 & \textbf{0.6566} & \textbf{0.3102}  \\
        
        $\ke{UP^{(2)}}$  & 0.0000 & 0.0000 & 0.0057 & 0.0059 &  0.0000 & 0.0052 & 0.0077  & \textbf{0.4896} & \textbf{0.1987} & \textbf{0.2643}  \\        

        \hline        
        \hline
    \end{tabular}
    \caption{The absolute squares of the coefficients $|c_{i,j}|^2$ of all polaritonic states for the case of two resonantly coupled  \ce{HF} molecules. The coefficients of the main contributions are highlighted in bold. The values are obtained using the full \gls{cbohf} energy expectation values, a cavity frequency $\omega_{c} = \omega_{1} = \SI{4281}{\per\centi\meter}$ and a coupling strength $\lambda_c$ of \SI{0.03}{\au}.}
    \label{tab:c2_hf2}
\end{table}

\begin{table}[h]
    \centering
    \begin{tabular}{c|c c c c c c}
        & $\ke{0,0}$ & $\ke{0,1}$ & $\ke{1,0}$ & $\ke{0,2}$ & $\ke{1,1}$ & $\ke{2,0}$  \\
        \hline
        $\ke{g}$  &   \textbf{0.9942} & 0.0057  &  0.0000 & 0.0000  & 0.0000   &  0.0000   \\
        \hline           
        $\ke{p^{(1)}}$  &  0.0057  & \textbf{0.9706} & 0.0124 & 0.0112 & 0.0000  & 0.0000     \\
        $\ke{e}$  & 0.0000 & 0.0122 & \textbf{0.9815} & 0.0001 & 0.0060 &  0.0000 \\
        \hline           
        $\ke{p^{(2)}}$  & 0.0000 & 0.0111 &  0.0002 & \textbf{0.9463} & 0.0248 & 0.0006 \\  
        $\ke{LP^{(2)}}$  & 0.0000 & 0.0002 & 0.0030 & 0.0162 & \textbf{0.4649}  & \textbf{0.5062}   \\
        $\ke{UP^{(2)}}$  & 0.0000  & 0.0000 & 0.0030 & 0.0090 & \textbf{0.4920} & \textbf{0.4866}   \\   
        \hline        
        \hline
    \end{tabular}
    \caption{The absolute squares of the coefficients $|c_{i,j}|^2$ of all polaritonic states for the case of a single resonantly coupled  \ce{HF} molecule. The coefficients of the main contributions are highlighted in bold. The values are obtained using the full \gls{cbohf} energy expectation values, a cavity frequency $\omega_{c} = \omega_{2} = \SI{4108}{\per\centi\meter}$ and a coupling strength $\lambda_c$ of \SI{0.03}{\au}.}
    \label{tab:c2_hf1_w2}
\end{table}

\begin{table}[h]
    \centering
    \begin{tabular}{c|c c c c c c c c c c}
        & $\ke{00,0}$ & $\ke{00,1}$ & $\ke{01,0}$ & $\ke{10,0}$ & $\ke{00,2}$ & $\ke{01,1}$ & $\ke{10,1}$ & $\ke{02,0}$  & $\ke{20,0}$   & $\ke{11,0}$ \\
        \hline

        $\ke{g}$  & \textbf{0.9771} &  0.0226 & 0.0000 & 0.0000 & 0.0003 & 0.0000 & 0.0000 & 0.0000 & 0.0000 & 0.0000 \\
         \hline     

        $\ke{p^{(1)}}$  & 0.0222 & \textbf{0.9104} & 0.0000 & 0.0229 & 0.0431 & 0.0003 & 0.0003 & 0.0000 & 0.0000 & 0.0000 \\
         
        $\ke{d_1}$ & 0.0000 & 0.0000 & \textbf{0.9763} & 0.0000 & 0.0000 & 0.0133 & 0.0101 & 0.0000 &  0.0000 & 0.0000 \\

        $\ke{e}$
        & 0.0004 & 0.0220 & 0.0000 & \textbf{0.9534} & 0.0010 & 0.0098 & 0.0129  & 0.0000 & 0.0000 & 0.0000 \\
        \hline   

        $\ke{p^{(2)}}$  & 0.0002 & 0.0424 &  0.0000 & 0.0011 & \textbf{0.8459} & 0.0188 & 0.0248  & 0.0005 & 0.0006 & 0.0003 \\  
        
         $\ke{d_2}$   & 0.0000 & 0.0000 &  0.0132 & 0.0000 & 0.0000 &  \textbf{0.2910} & \textbf{0.2205}  & \textbf{0.2430} & \textbf{0.1955}  & 0.0000 \\

        $\ke{LP^{(2)}}$   & 0.0000 & 0.0013 & 0.0000 & 0.0115 &  0.0284 & \textbf{0.1906} & \textbf{0.2516} & \textbf{0.2099} & \textbf{0.2609}  & 0.0092 \\

       $\ke{d_3}$  & 0.0000 &  0.0000 &  0.0102 & 0.0000 &  0.0000 & \textbf{0.2383} & \textbf{0.1805} & \textbf{0.2976}  & \textbf{0.2395} & 0.0000 \\     
 
        $\ke{UP^{(2)}}$ & 0.0000 & 0.0006 & 0.0000 & 0.0103 &  0.0146 & \textbf{0.1811} & \textbf{0.2390} & \textbf{0.2243} & \textbf{0.2787}  & 0.0166 \\     

       $\ke{f_2}$  & 0.0000 &  0.0000 &  0.0000 & 0.0005 &  0.0003 & 0.0104 & 0.0137 & 0.0003 & 0.0004 & \textbf{0.9495}\\    
        \hline        
        \hline
    \end{tabular}
    \caption{The absolute squares of the coefficients $|c_{i,j}|^2$ of all polaritonic states for the case of two resonantly coupled  \ce{HF} molecules. The coefficients of the main contributions are highlighted in bold. The values are obtained using the full \gls{cbohf} energy expectation values, a cavity frequency $\omega_{c} = \omega_{2} = \SI{4108}{\per\centi\meter}$ and a coupling strength $\lambda_c$ of \SI{0.03}{\au}.}
    \label{tab:c2_hf2_w2}
\end{table}

\clearpage

\section{DQC spectra for a cavity resonant with the first hot transition of \ce{HF}}

In this section, we briefly discuss the formed vibrational polaritons and the corresponding \gls{dqc} spectra when the cavity mode is resonant with the first hot transition of the \ce{HF} molecule, $\omega_c = \omega_2 = \SI{4108}{\per\centi\meter}$.

\begin{figure}[htb!]
     \centering
    \includegraphics[width=0.60\textwidth]{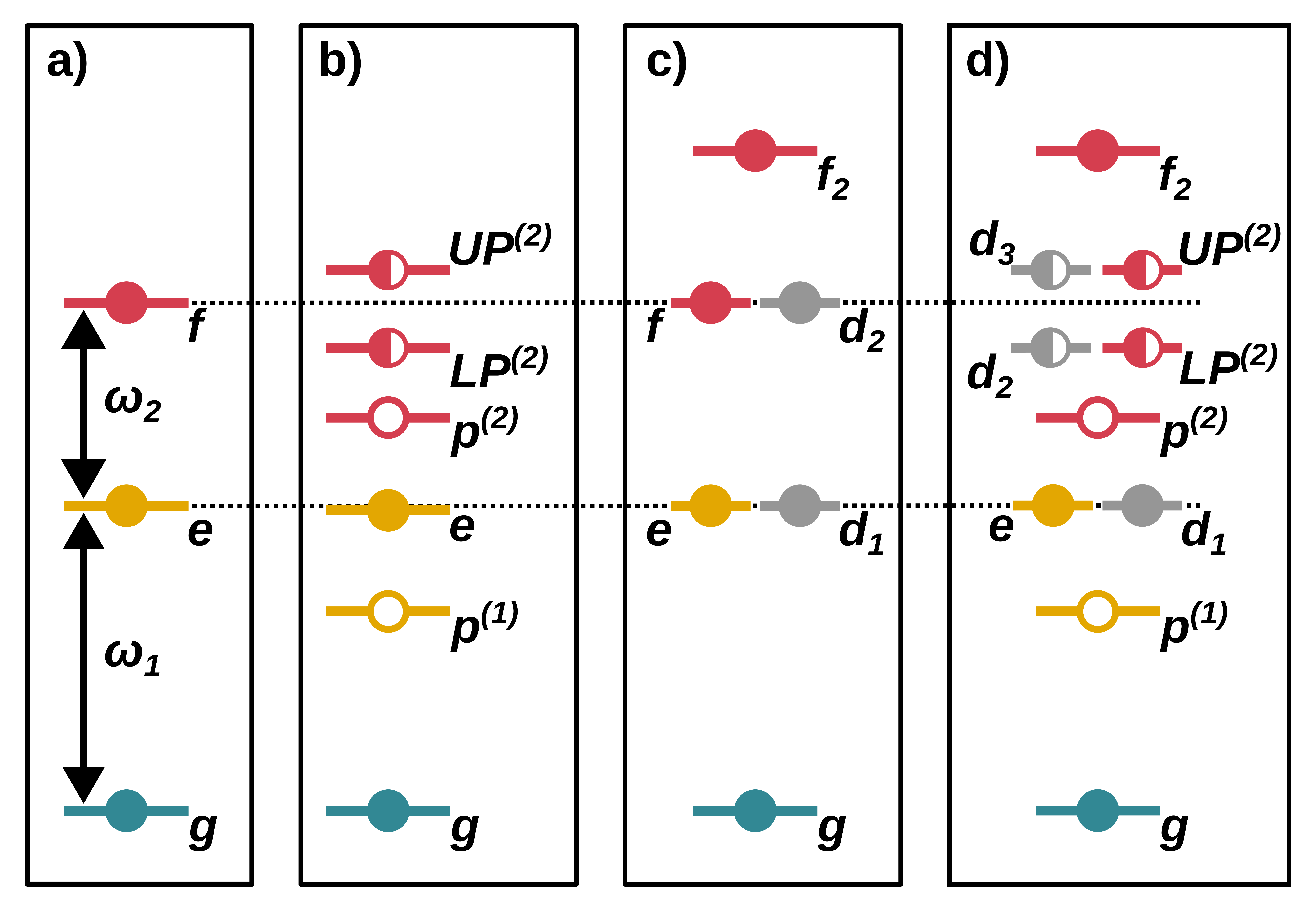}
    \caption{Schematic energy level diagrams for a single \ce{HF} molecule and a pair of \ce{HF} molecules without a cavity in a) and c) and resonantly coupled with a single cavity mode in b) and d). The cavity frequency $\omega_c$ is resonant with $\omega_2$. The ground state is colored green, the single excited state is colored yellow, and the double excited states are colored red. The optically dark states originating from the ground state to the single excited manifold and from the single excited manifold to the double excited manifold are shown in gray. Full circles indicate states of predominantly matter character, empty circles indicate states of predominantly photonic character, and half-filled circles indicate states with a mixed matter and photon contribution.} 
 \label{fig:4108_levels}
\end{figure}

Figure~\ref{fig:4108_levels} shows the corresponding schematic energy level diagrams for a single \ce{HF} molecule and a pair of \ce{HF} molecules without a cavity in a) and c) and resonantly coupled with a single cavity mode in b) and d). 
Apart from the expected energetic differences between the results obtained with the three different energy expectation values, the schematic energy-level diagrams shown are the same in all three cases.
A analysis of the discussed polaritonic states for both the single-molecule case and the two-molecule case in terms of uncoupled bare states $\ke{k,n}$ can be found in section~S1 TABLE~\ref{tab:c2_hf1_w2} and TABLE~\ref{tab:c2_hf2_w2}. 

Since the uncoupled energy levels are discussed in the manuscript section~III~A, we restrict the discussion to the two cavity-coupled cases shown in Figure~\ref{fig:4108_levels}~b) and d). 
Within the first excitation manifold (color coded yellow), the cavity mode is out of resonance $\omega_c = \omega_1 - \Delta$ due to the anharmonicity of \ce{HF}, and consequently the molecular state $e$ (and $d_1$ in the case of two molecules) is almost unchanged compared to the field-free situation. 
However, because of the remaining weak off-resonance interaction, the photonic $p^{(1)}$ state (bare state $\ke{0,1}$) still has a small molecular contribution.

The expected formation of the \gls{lp} state and the \gls{up} state is observed only in the second excitation manifold, since the cavity is resonant with the first hot transition.
As shown in Figure~\ref{fig:4108_levels}~b) Figure~\ref{fig:4108_levels}~d), the corresponding $UP^{(2)}$ state and the $LP^{(2)}$ state are separated in both cases by a Rabi splitting $\Omega_R^{(2)}$ of about \SI{60}{\per\centi\meter}. 
In the case of a single \ce{HF} molecule, these polaritonic states are formed by a linear combination of the two bare states $\ke{1,1}$ and $\ke{2,0}$. 
The remaining state $p^{(2)}$ is almost fully photonic (bare state $\ke{0,2}$) and has only a small molecular contribution. 
For two \ce{HF} molecules coupled to the cavity mode, in addition to the two polaritonic states $UP^{(2)}$ and $LP^{(2)}$, two energetically degenerate dark states $d_2$ and $d_3$ are present. 
These dark states are also formed by the four bare states $\ke{01,1}$, $\ke{10,1}$, $\ke{02,0}$, and $\ke{20,0}$ similarly to $UP^{(2)}$ and $LP^{(2)}$, but the leading components are excitations of the dark antisymmetric stretching modes ($\ke{011}$ and $\ke{02,1}$).  
The remaining doubly excited state $f_2$ is $\Delta$ higher in energy than the average of the $UP^{(2)}$ state and the $LP^{(2)}$ state and is formed by a single simultaneous excitation of both stretching modes ($\ke{11}$).

\begin{figure}[ht!]
     \centering
    \includegraphics[width=0.9\textwidth]{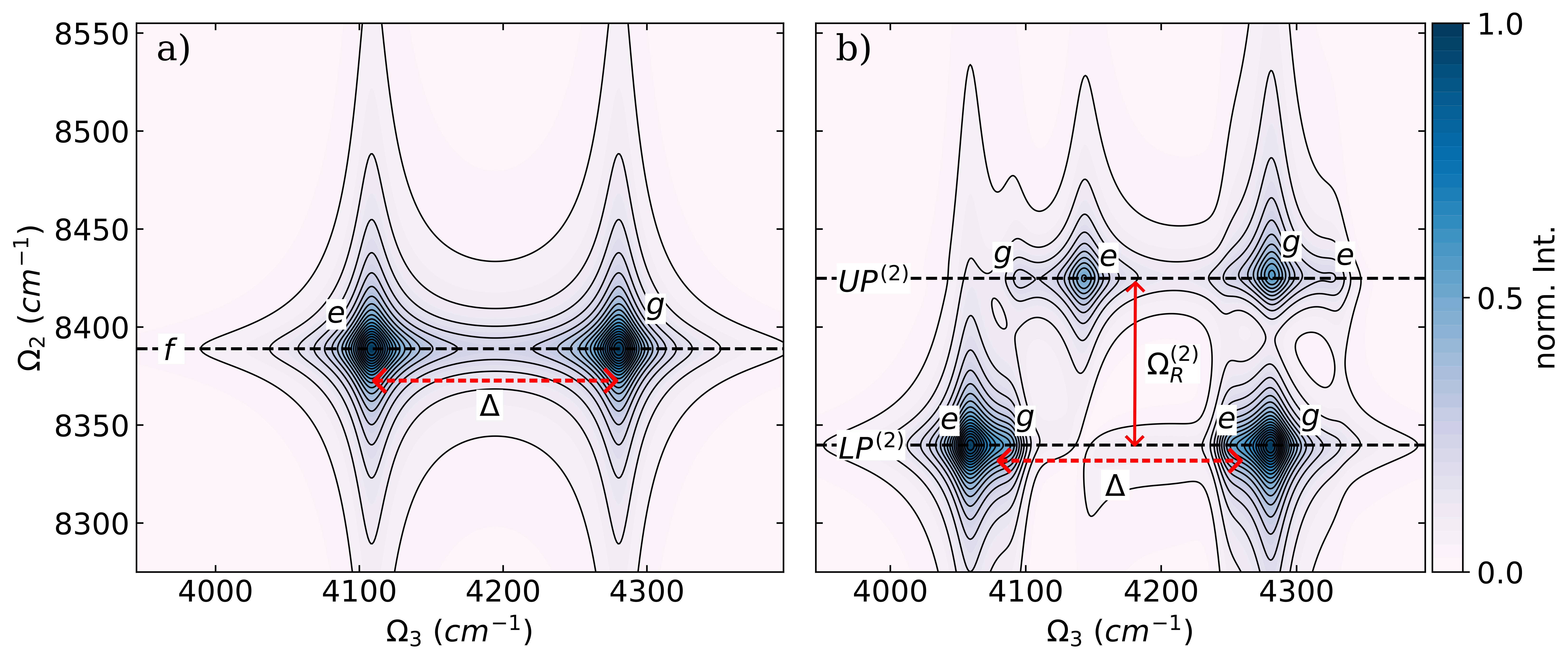}
    \caption{Absolute value of the normalized \gls{dqc} spectra of a single \ce{HF} molecule a) without a cavity and b) coupled to a single cavity mode with $\omega_{c} = \SI{4108}{\per\centi\meter} $. The coupling strength $\lambda_c$ is \SI{0.03}{\au} and the dephasing $\gamma$ is \SI{10}{\per\centi\meter}. The black horizontal dashed lines mark the energy of the final states, and all signals are labeled $e$ and $g$, indicating that the initial state is the ground state or an intermediate state. The red lines with arrows highlight relevant energy differences. } 
\label{fig:1hf_dqc}
\end{figure}

The corresponding normalized absolute values of the \gls{dqc} spectra for one \ce{HF} molecule are shown in Figure~\ref{fig:1hf_dqc}.
For the coupled single-molecule case shown in Figure~\ref{fig:1hf_dqc}~b), the main difference from situation $\omega_c =\omega_1$ is the presence of only two resonances on the $\Omega_2$ axis associated with the two hybrid states $UP^{(2)}$ and $LP^{(2)}$ separated by the Rabi splitting $\Omega_R^{(2)}$. The peaks corresponding to the final state $LP^{(2)}$ are more intense, but for both resonances two main peaks and two weak side bands are observed. The main peaks are transitions into (around \SI{4300}{\per\centi\meter}) and out of (around \SI{4050}{\per\centi\meter}) the molecular state $e$, while the weak side bands are transitions involving the mostly photonic state $p^(1)$.

\begin{figure}
     \centering
    \includegraphics[width=0.9\textwidth]{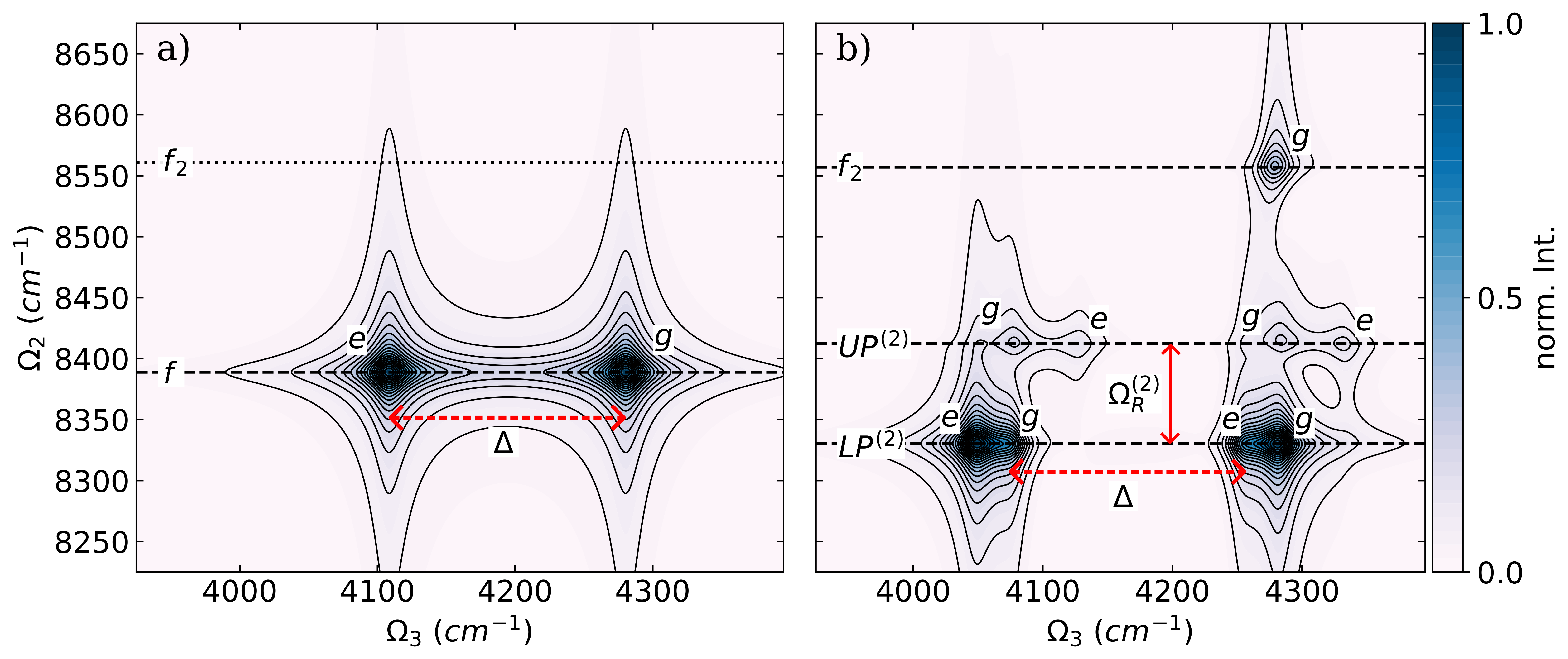}
    \caption{Absolute value of the normalized \gls{dqc} spectra of two parallel oriented \ce{HF} molecules a) without a cavity and b) coupled to a single cavity mode with $\omega_{c} = \SI{4108}{\per\centi\meter} $. The coupling strength $\lambda_0$ is \SI{0.03}{\au} and the dephasing $\gamma$ is \SI{10}{\per\centi\meter}. The black horizontal dashed lines mark the energy of the final states, and all signals are labeled $e$ and $g$, indicating that the initial state is the ground state or an intermediate state. The red lines with arrows highlight relevant energy differences.} 
\label{fig:2hf_dqc}
\end{figure}

The corresponding normalized absolute values of the \gls{dqc} spectra for two \ce{HF} molecules are shown in Figure~\ref{fig:2hf_dqc}. 
The main difference in the \gls{dqc} spectra for the two coupled \ce{HF} molecules shown in Figure~\ref{fig:2hf_dqc}~b) compared to the case of a single molecule is the presence of only three resonances. In addition to the two resonances corresponding to the final states $UP^{(2)}$ and $LP^{(2)}$, there is a third resonance on the $\Omega_2$ axis associated with the final state $f_{2}$. Since this state is formed by the simultaneous single excitation of stretching modes ($\ke{11}$), it cannot exist in the case of a single molecule and is only visible when the two molecules interact with the cavity mode.
Without the cavity interaction, $f_{2}$ is not visible in the \gls{dqc} spectra due to a cancellation of the Liouville diagrams, since the corresponding energy splitting is harmonic in nature.
In the cavity, however, the harmonic energy splitting is broken by the light-matter interaction, which introduces additional anharmonicity into the system. 
The same situation is observed for the formation of the state $MP^{(2)}$ in the case of $\omega_c = \omega_1$ discussed in the main manuscript. 

\begin{figure}
     \centering
    \includegraphics[width=0.9\textwidth]{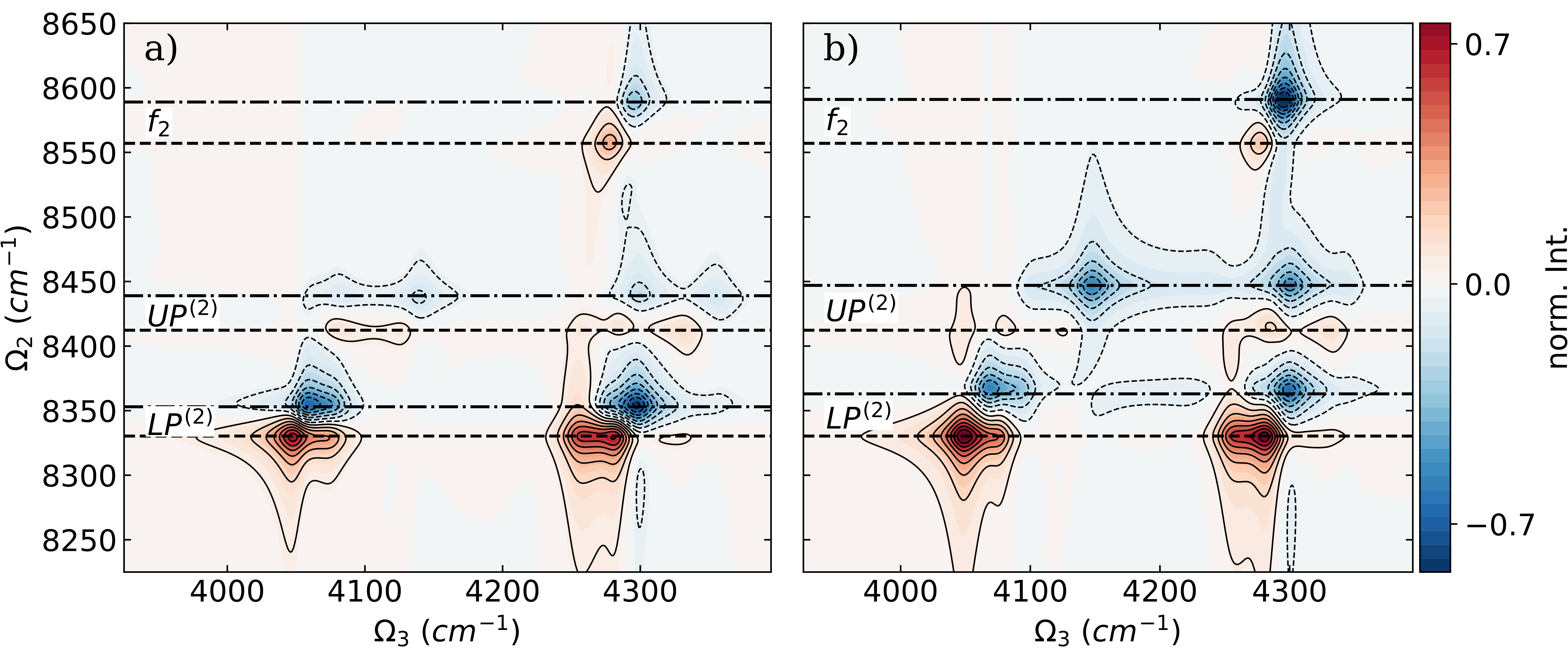}
    \caption{Difference $\Delta \mathcal S $ of the \gls{dqc} signal of two parallel \ce{HF} molecule coupled to a photon mode with $\omega_{c} = \SI{4108}{\per\centi\meter}$ between a)  full \gls{cbohf} and linear \gls{cbohf}  and b) full \gls{cbohf} and \gls{etc}. Individual \gls{dqc} spectra are normalized, and the absolute value is used to calculate the difference. The coupling strength $\lambda_c$ is \SI{0.03}{\au} for both frequencies and the dephasing $\gamma$ is \SI{10}{\per\centi\meter}.  The energies of the final states are marked with black dashed lines for the full \gls{cbohf} case and with dashed dotted lines for the linear \gls{cbohf} and \gls{etc} cases, respectively.} 
\label{fig:2hf_dqc_delta}
\end{figure}

In Fig.~\ref{fig:2hf_dqc_delta} the comparison between the \gls{dqc} signals based on the full \gls{cbohf} surface, the linear \gls{cbohf} surface, and the \gls{etc} surfaces for the two-molecule case is shown as difference spectra $\Delta \mathcal S $. 
In both difference spectra the peaks corresponding to the final state $LP^{(2)}$ are the most affected. 
These peaks are red-shifted by approximately \qtyrange{20}{30} {\per\centi\meter} in $\Omega_2$ and approximately \qtyrange{10}{20}{\per\centi\meter} in $\Omega_3$ when the \gls{dse} contribution is included, see Fig.~\ref{fig:2hf_dqc_delta}~a), or the \gls{scf} procedure is performed, see Fig.~\ref{fig:2hf_dqc_delta}~b)).  
A similar but weaker red shift is observed for the peaks associated with the final state $UP^{(2)}$.
The \gls{dqc} signal corresponding to the final state $f_{2}$ is also red-shifted when the \gls{dse} contribution is included, while the intensity is more or less unchanged. However, without the \gls{scf} treatment, the corresponding signal is not only shifted in both $\Omega_2$ and $\Omega_3$, but also the intensity is much higher. This result clearly indicates the sensitivity of this $f_2$ state to the way the cavity is described.
\clearpage
\section{Real and imaginary parts of the DQC spectra}

\begin{figure}[!htb]
     \centering
    \includegraphics[width=0.9\textwidth]{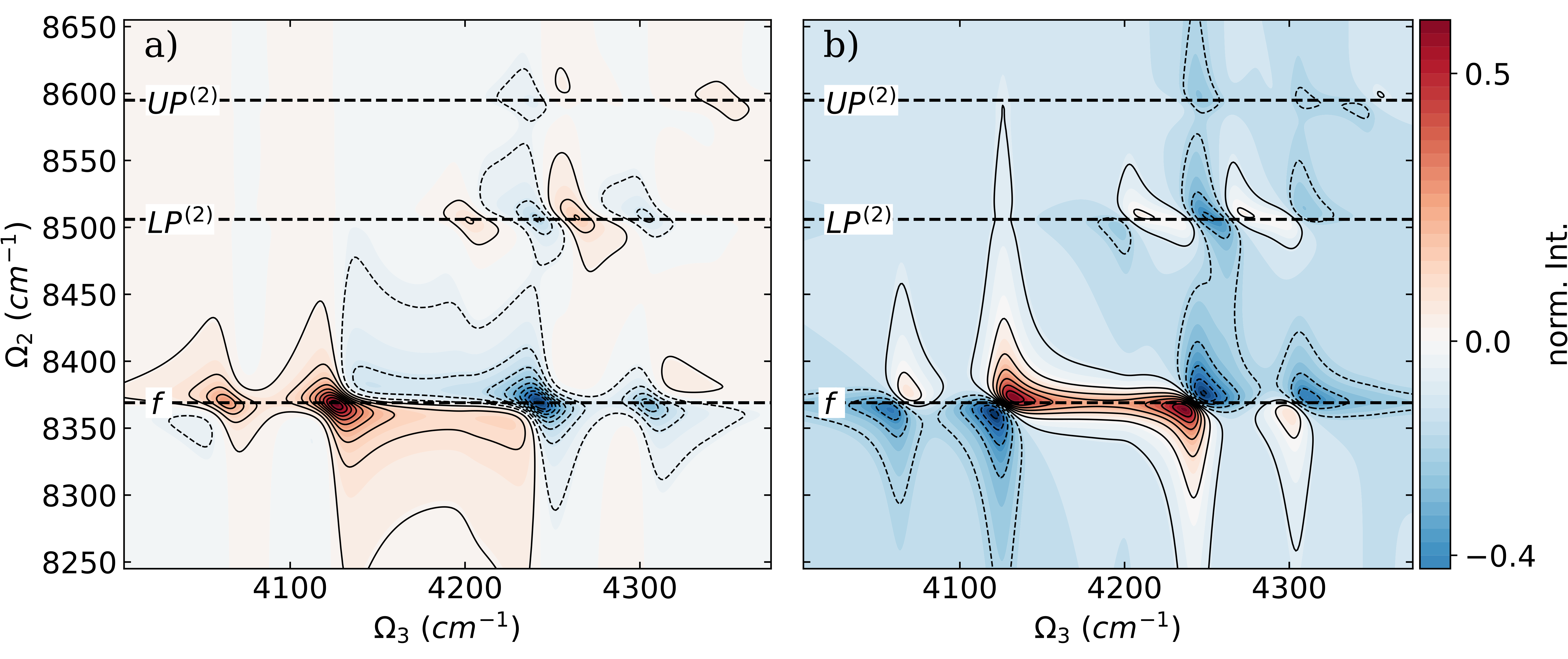}
    \caption{ a) Real ($Re$) part and b) imaginary ($Im$) part of the \gls{dqc} signal of a single \ce{HF} molecule coupled to a photon mode with $\omega_{c} = \SI{4281}{\per\centi\meter}$. Both are normalized with respect to the absolute value of the \gls{dqc} signal. The coupling strength $\lambda_c$ is \SI{0.03}{\au} and the dephasing $\gamma$ is \SI{10}{\per\centi\meter}.  The black horizontal dashed lines mark the energy of the final states.} 
\label{fig:1hf_dqc_part_w1}
\end{figure}

\begin{figure}
     \centering
    \includegraphics[width=0.9\textwidth]{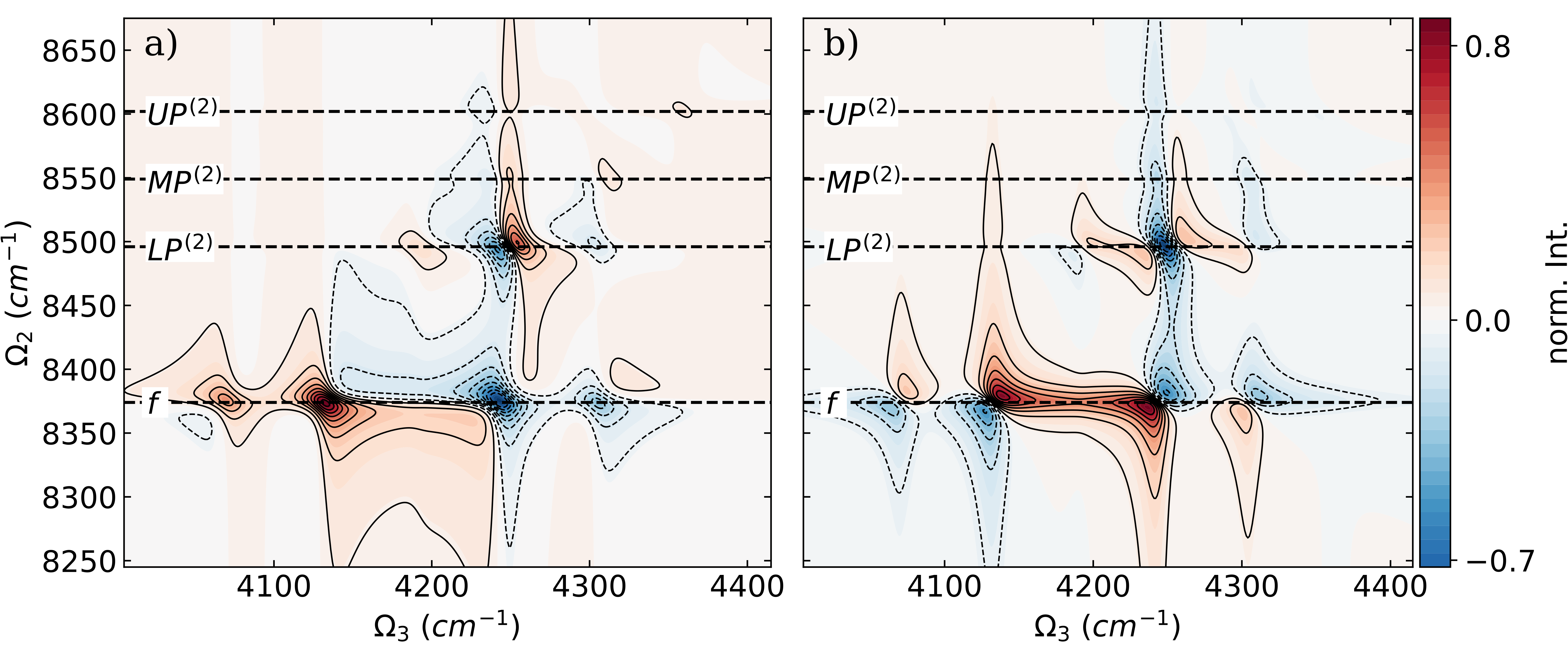}
    \caption{ a) Real ($Re$) part and b) imaginary ($Im$) part of the \gls{dqc} signal of two parallel \ce{HF} molecules coupled to a photon mode with $\omega_{c} = \SI{4281}{\per\centi\meter}$. Both are normalized with respect to the absolute value of the \gls{dqc} signal. The coupling strength $\lambda_c$ is \SI{0.03}{\au} and the dephasing $\gamma$ is \SI{10}{\per\centi\meter}.  The black horizontal dashed lines mark the energy of the final states.} 
\label{fig:2hf_dqc_part_w1}
\end{figure}

\begin{figure}
     \centering
    \includegraphics[width=0.9\textwidth]{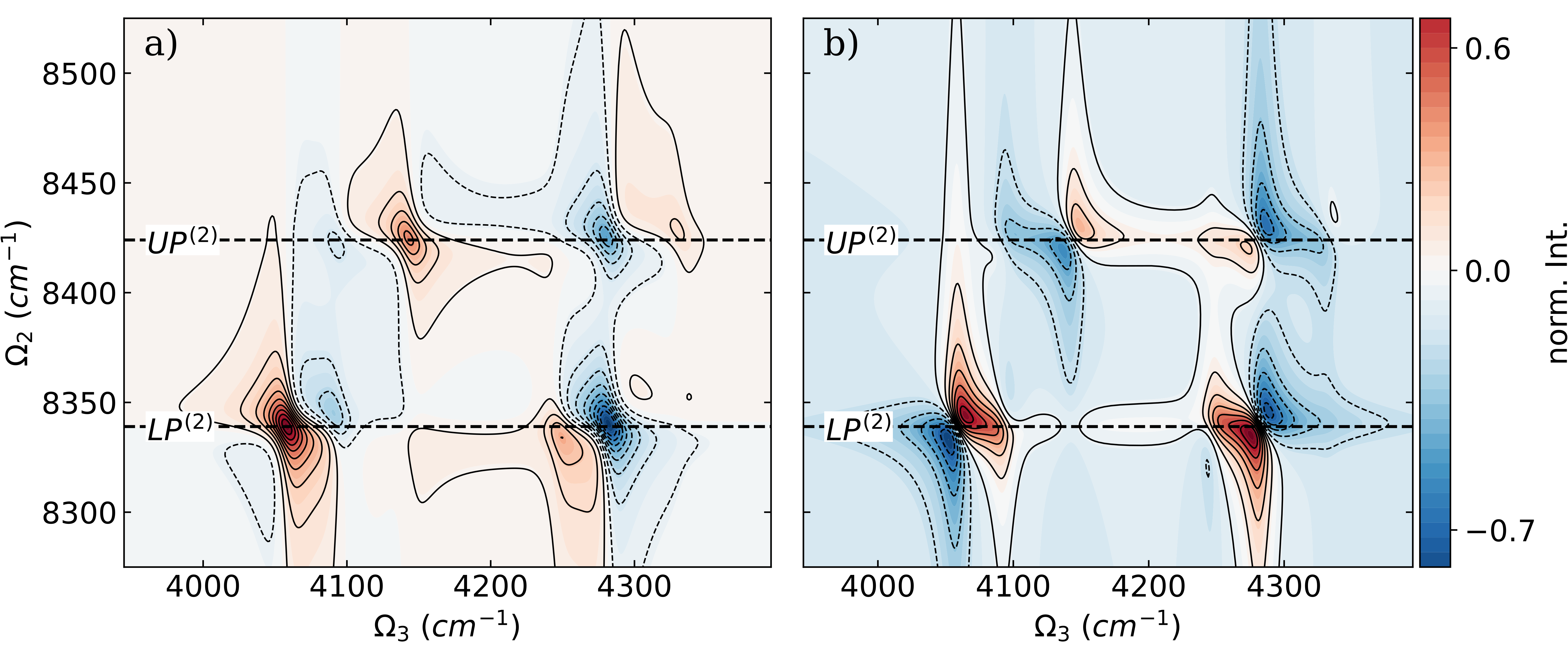}
    \caption{ a) Real ($Re$) part and b) imaginary ($Im$) part of the \gls{dqc} signal of a single \ce{HF} molecule coupled to a photon mode with $\omega_{c} = \SI{4108}{\per\centi\meter}$. Both are normalized with respect to the absolute value of the \gls{dqc} signal. The coupling strength $\lambda_c$ is \SI{0.03}{\au} and the dephasing $\gamma$ is \SI{10}{\per\centi\meter}.  The black horizontal dashed lines mark the energy of the final states.} 
\label{fig:1hf_dqc_part_w2}
\end{figure}

\begin{figure}
     \centering
    \includegraphics[width=0.9\textwidth]{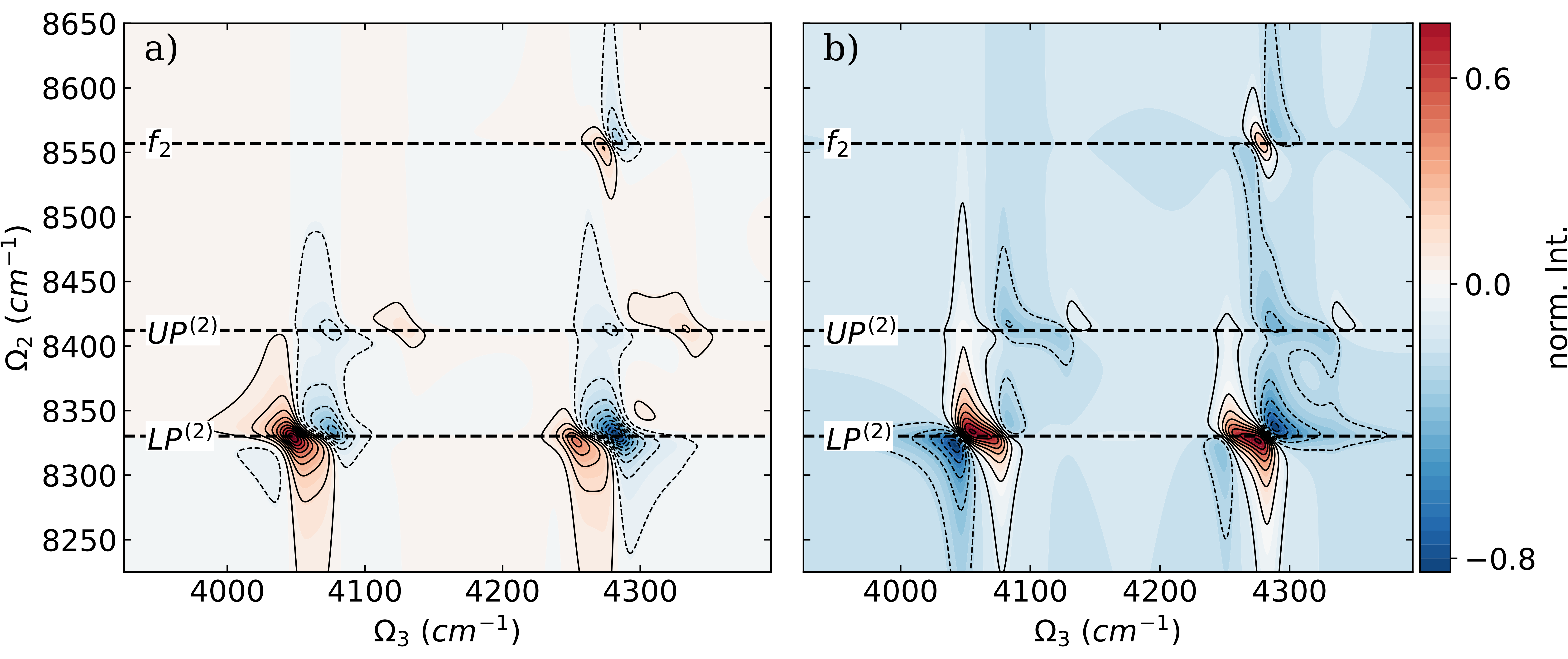}
    \caption{a) Real ($Re$) part and b) imaginary ($Im$) part of the \gls{dqc} signal of two parallel \ce{HF} molecules coupled to a photon mode with $\omega_{c} = \SI{4108}{\per\centi\meter}$. Both are normalized with respect to the absolute value of the \gls{dqc} signal. The coupling strength $\lambda_c$ is \SI{0.03}{\au} and the dephasing $\gamma$ is \SI{10}{\per\centi\meter}.  The black horizontal dashed lines mark the energy of the final states.} 
\label{fig:2hf_dqc_part_w2}
\end{figure}

The real ($Re$) and imaginary ($Im$) parts of all discussed \gls{dqc} spectra are shown in Fig.~\ref{fig:1hf_dqc_part_w1}, Fig.~\ref{fig:2hf_dqc_part_w1} ,Fig.~\ref{fig:1hf_dqc_part_w2}, and Fig.~\ref{fig:2hf_dqc_part_w2} using the full \gls{cbohf} energy expectation values and a coupling strength $\lambda_c$ of \SI{0.03}{\au}.

%\bibliography{lit.bib}